\documentclass[pre,superscriptaddress,twocolumn,floatfix]{revtex4-2}

\usepackage{microtype}
\usepackage[utf8]{inputenx}
\usepackage{physics}                            
\usepackage{bm}									
\usepackage{enumerate}							
\usepackage{amsthm}								
\usepackage{comment}							
\usepackage[caption=false]{subfig}
\usepackage{amssymb}							
\usepackage{t1enc}								
\usepackage{mathtools}							
\usepackage{tensor}								
\usepackage{wrapfig}							
\usepackage{listings}							
\usepackage{xcolor}								
\usepackage{hyperref}
\hypersetup{colorlinks=true}
\usepackage[capitalize]{cleveref}               
\usepackage{graphicx,color}
\usepackage{amsfonts,amsmath}
\usepackage{xstring}                            
\usepackage{tikz}
\usepackage{lettrine}
\usetikzlibrary{positioning}

\newcommand{\cor}[1]{\mathcal{#1}}									
\newcommand{\T}[1]{\text{#1}}										
\newcommand{\eg}{{\em e.g.}}
\newcommand{\ie}{{\em i.e.}}
\newcommand{\n}{\nonumber}
\newcommand{\ccite}[1]{\IfSubStr{#1}{,}{Refs.~}{Ref.~}\cite{#1}}

\begin{document}

\title{Collective response to local perturbations:\\how to evade threats without losing coherence}

\author{Emanuele Loffredo}
\email{emanuele.loffredo@phys.ens.fr}
\affiliation{Laboratory of Physics of the École Normale Supérieure, CNRS UMR 8023 and PSL Research, 24 rue Lhomond, 75231 Paris Cedex 05, France}
\author{Davide Venturelli}
\email{davide.venturelli@sissa.it}
\affiliation{SISSA -- International School for Advanced Studies and INFN, via Bonomea 265, 34136 Trieste, Italy}
\author{Irene Giardina}
\affiliation{Dipartimento di Fisica, Universit\a`{a} Sapienza, 00185 Rome, Italy}
\affiliation{Istituto Sistemi Complessi, Consiglio Nazionale delle Ricerche, UOS Sapienza, 00185 Rome, Italy}
\affiliation{INFN, Unit\a`{a} di Roma 1, 00185 Rome, Italy}

\begin{abstract}
Living groups move in complex environments and are constantly subject to external stimuli, predatory attacks and disturbances. An efficient response to such perturbations is vital to maintain the group's coherence and cohesion. Perturbations are often local, \textit{i.e.} they are initially perceived only by few individuals in the group, but can elicit a global response. This is the case of starling flocks, that can turn very quickly to evade predators. In this paper, we investigate the conditions under which a global change of direction can occur upon local perturbations. Using minimal models of self-propelled particles, we show that  a collective directional response occurs on timescales that grow with the system size and it is, therefore, a finite-size effect. The larger the group is, the longer it will take to turn. We also show that global coherent turns can only take place if i) the mechanism for information propagation is efficient enough to transmit the local reaction undamped through the whole group; and if ii) motility is not too strong, to avoid 
that
the perturbed individual
leaves
the group before the turn is complete. No compliance with such conditions results in the group's fragmentation or in a non-efficient response.
\end{abstract}

\maketitle

\section{Introduction}

%
%


Collective behaviour in living aggregations often has a strong anti-predatory function. An efficient group response is crucial to maintain global coherence and cohesion in spite of attacks and disturbances \cite{LIMA_1995,Krause2002living,Couzin_2011,Rosenthal_2015,Handegard2012,HerbertRead2015,Calovi_2015,Klamser2021}.
To this end, many groups exhibit collective directional changes in very short times, often triggered by local perturbation events. A beautiful example is the one of bird flocks 
\cite{information_transfer}, 
where collective turns start from a single individual, and the directional information is passed through the group so quickly that the whole flock performs the turn retaining its structure and without attenuation. 
From the perspective of statistical physics, this behaviour is somewhat unusual. Indeed, when considering physical systems with long-range directional order -- either at equilibrium or active -- perturbing locally the system does not in general change its ordered state, which is another way of saying that the order is stable (\ie{} ergodicity is broken in the thermodynamic limit). For instance, if we consider a ferromagnet and apply a magnetic field on a single site of the lattice, the magnetization will remain unaltered. 
What marks the difference with respect to animal groups is, of course, their size. Living groups might be large, comprising hundreds or thousands of individuals, but they are very far from the order Avogadro numbers that characterize physical systems. Perturbations which would be irrelevant in the thermodynamic limit might in fact change the state of a finite group on observational timescales, if its size is small enough. Response to local perturbations is thus, intrinsically, a finite-size effect. 
The kind of response exhibited by the system, and the effective timescale for collective adaptation, also depend on the mechanism of information propagation. In flocks, a linear dispersion law has been observed 
\cite{information_transfer}, 
suggesting that second order dynamics for the birds flight directions should be at play \cite{flocking_and_turning,cavagna2015silent}. 
Other animal groups display instead less efficient information transfer. 
For instance, experiments on fish schools \cite{Calovi_2015,Rosenthal_2015,Couzin_2011}
show that groups exhibit evasion manoeuvres arising from induced alarming stimuli. However -- contrary to bird flocks -- the speed of information propagation decelerates over time, and thus only part of the group follows the initiator of the evasion event, resulting in a distribution of behavioural cascades. In this case, a dissipative dynamics rules individual directional motion, and the local connectivity of the interaction network might dramatically affect the extension of the collective response \cite{Rosenthal_2015}.

These examples indicate that several factors might contribute to the response behaviour of finite groups, and it is not always clear how to disentangle one factor from the other. In this work we perform a systematic study to explore the interplay between size, dynamical rules, motility and boundary conditions in determining the response of the system to local directional perturbations. The framework of our study is the one of self-propelled models of collective motion, where the minimal number of parameters allows to exhaustively explore the model's space. 

This paper is organized as follows. In \cref{par:model} we introduce the model discussed in our theoretical and numerical analysis: the Inertial Spin Model (ISM \cite{flocking_and_turning}). This model is a generalisation of the Vicsek model \cite{vicsek} comprising both inertial and dissipative terms in the dynamical equation for the velocities. In the underdamped limit it reproduces the linear dispersion law observed in flocks of birds (for which it was originally proposed), while in the overdamped limit it
reduces to the Vicsek model. It therefore represents an ideal playground to investigate different dynamical regimes potentially relevant for different behaviours. In the following sections we consider progressively all the factors that might affect the response behaviour. In \cref{par:eq_onlattice} we start by looking at the ISM with a fixed interaction network. In this case it is possible to derive analytically the equations describing the evolution of the system, when a local perturbation is applied to a single individual. 
This allows to clearly understand what is the role of the system's size, and how a collective turn manifests itself in the different dynamical regimes of the model. In \cref{par:onlattice} we perform numerical simulations that quantitatively confirm our analytical results. Next, in \cref{par:offlattice} we consider the off-lattice, active version of the model, and elucidate how motility affects the fixed-network scenario. Finally, in \cref{par:openBC} we discuss what happens when open boundary conditions are considered, and we show how and when the mechanisms studied in the previous sections can lead to a fragmentation of the group.

\section{The modelling framework} 
\label{par:model}
Collective behaviour in animal groups has been described with a variety of approaches, either considering evolution rules for the individuals in the group \cite{Vicsek_review,couzin_review}, 
or by using coarse-grained equations for mesoscopic fields \cite{toner_review,marchetti_review,ramaswamy_review}. In this study, we work in the context of self-propelled particle (SPP) models, where the aggregation is modelled in a minimal way as a collection of particles with fixed
activity interacting with each other via alignment/imitation rules.

\subsection{The Vicsek model}
\label{par:vicsek}
The most famous among SPP models of collective motion is the Vicsek model (VM \cite{vicsek}), which we now present in its continuous-time version. Consider a system made of $N$ point-like active particles, each labeled by an index $i$, which move in a $d$-dimensional space following the equations of motion
\begin{align}
    \dv{\vb{r}_i}{t} &= \vb{v}_i \, , \label{eq:pos} \\
    \eta \dv{\vb{v}_i}{t} &=  - \pdv{\cor{U}}{\vb{v}_i}+ \lambda_i \vb{v}_i + \bm{\xi}_i \, .
    \label{eq:vicsek_v}
\end{align}
Each particle position $\vb{r}_i\in \mathbb{R}^d$ evolves deterministically according to its corresponding velocity vector $\vb{v}_i \in \mathbb{R}^{d_v} $ of fixed modulus $\abs{\vb{v}_i}=v_0$, while the latter undergoes a Langevin overdamped dynamics with a ``social'' alignment force given by
\begin{align}
\cor{U}(\{\vb{v}_i\}) & =  -\frac{J}{2v_0^2}\sum_{ij} n_{ij} \vb{v}_i \vdot \vb{v}_j \, , \label{eq:heisenberg}\\ 
 - \pdv{\cor{U}}{\vb{v}_i}& = \frac{J}{v_0^2}\sum_{j} n_{ij} \vb{v}_j  \ .
\end{align}
We recognize in \cref{eq:heisenberg} the Hamiltonian of the Heisenberg model, with the coefficient $J>0$ setting the strength of the alignment interactions, and where the connectivity matrix $n_{ij}$ specifies the interacting neighbours. However, in contrast to the standard Heisenberg model, here the connectivity matrix itself evolves in time through the particle positions, \ie{} $n_{ij}=n_{ij}(\left\lbrace \vb{r}_i(t) \right\rbrace_{i=1}^N )$. Among the possible evolution rules, we will adopt in the following the paradigm of \textit{metric} interactions: each particle interacts with those lying within a fixed interaction radius $r_c$, \ie{} $n_{ij}(t)=\Theta(r_c- \abs{\vb{r}_i(t)-\vb{r}_j(t)} )$.

Featuring in \cref{eq:vicsek_v} are a viscosity coefficient $\eta$ and a white Gaussian noise $\bm{\xi}_i$, which are linked by the Einstein-like relation
\begin{equation}
    \expval{ \bm{\xi}_i(t) \cdot \bm{\xi}_j (t') } = 2d\eta T \delta_{ij} \delta(t-t') \, ,
    \label{eq:noise_corr}
\end{equation}
where the \textit{temperature} $T$ sets the strength of the fluctuations. Finally, the parameter $\lambda_i$ is a Lagrange multiplier which can be used to enforce the fixed \textit{speed} condition $d\abs{\vb{v}_i}^2/dt=0$ \cite{Review_2018}.

Due to the time dependence in $n_{ij}(t)$, it is known that the Vicsek model admits, even in $d\leq 2$, a $T$-driven transition from an ordered phase (\textit{flock}) where the individual velocity vectors align and add up to a non-zero total velocity
\begin{equation}
\label{eq:order_param}
    \vb{V} \equiv \frac{1}{N} \sum_{i=1}^{N} \vb{v}_i \, ,
\end{equation}
to a disordered phase (\textit{swarm}) where the mean group velocity $\vb{V}$ is null. A convenient order parameter to describe the transition is the so-called polarization vector 
\begin{equation}
\label{eq:pol_vec}
    \vb{\Psi} \equiv \frac{1}{N} \sum_{i=1}^{N} \frac{\vb{v}_i}{|\vb{v}_i|} = \frac{\vb{V}}{v_0}\, ,
\end{equation}
and the associated (scalar) polarization $\Psi=|\vb{\Psi}| \in [0,1]$.

\subsection{The Inertial Spin Model}
The Vicsek model successfully describes the large-scale behaviour of several living and non-living active systems \cite{Vicsek_review, marchetti_review}. However, the VM is not appropriate to explain collective turns in flocks of birds. Experiments indeed show that, in turning flocks, the directional information travels obeying a linear dispersion law, with a propagation speed only depending on the degree of order in the system \cite{information_transfer}.
The Vicsek dynamics does not reproduce this behaviour. An intuition of why this occurs can be grasped by looking at \cref{eq:vicsek_v}: its structure is that of an overdamped Langevin equation for the velocity, and this kind of equations usually lead to a diffusive dispersion law \cite{zwanzig_book}. 
The simplest way to obtain linearly dispersive solutions is to reinstate a second order derivative in \cref{eq:vicsek_v}, and this is precisely what the Inertial Spin Model does \cite{flocking_and_turning}. As in standard second order equations, it is convenient to write the model as a system of first order equations, by introducing appropriate conjugate variables. The ISM then reads 
\begin{align}
    \dv{\vb{r}_i}{t} &= \vb{v}_i \, ,\\
    \dv{\vb{v}_i}{t} &= \frac{1}{\chi} \vb{s}_i \cp \vb{v}_i \, ,     \label{eq:ism_micro_v} \\
     \dv{\vb{s}_i}{t} &= \frac{\vb{v}_i}{v_0} \cp \left( 
      - \frac{1}{v_0}\pdv{\cor{U}}{\vb{v}_i} -\frac{\eta}{v_0} \dv{\vb{v}_i}{t} + \bm{\xi}_i \right) \, ,
    \label{eq:ism_micro}
\end{align}
with the noise correlations given in \cref{eq:noise_corr}. In contrast to the Vicsek model, in \cref{eq:ism_micro}  the force term does not act directly on the velocity, but rather on its derivative, which is expressed in terms of the new variable $\vb{s}_i$. The vectorial products enforce the constraint on the individual speeds. The particles can thus only change their directions: $\vb{s}_i$ therefore plays the role of an internal angular momentum, regulating the rotations of the individual velocity vectors, and this is why it is called a `spin'. The new parameter $\chi$ plays the role of a rotational \textit{inertia}. 

\subsection{The planar ISM}
Let us now specialize the Inertial Spin Model to the planar two-dimensional case, where velocities lie on a plane ($d_v=2$). This case is simpler to handle algebraically, and it gives a direct intuition of the physical and biological meaning of the spin and of inertial dynamics. Besides, it is also the relevant one for collective turns in flocks (as we will discuss later on), and for a variety of other biological groups. Generalization to the three-dimensional case is straightforward. 

In the planar case, we can write
\begin{equation}
    \vb{v}_i(t) \equiv v_0 \left( \cos\varphi_i(t)\, , \,  \sin\varphi_i(t) \right) \, ,
    \label{eq:2d_parametrization}
\end{equation}
where the phase ${\varphi}_i$ specifies the angle of the velocity vector with respect to a reference direction. In terms of the phases, \cref{eq:ism_micro_v} and \cref{eq:ism_micro} assume a much simpler form, namely
\begin{align}
    \dot{\varphi}_i &= \frac{s_i}{\chi} \, , \label{eq:ism_plan1}\\
    \dot{s}_i &= 
      - \pdv{\cor{U}}{{\varphi}_i}
    -\dfrac{\eta}{\chi} s_i + \xi_i \, .
    \label{eq:ism_plan2}
\end{align}
In these equations the spin vector reduces to one single component, whose value $s_i$ identifies the angular velocity of each particle. In the presence of a force, the spin acquires a non-zero value and, as a consequence, the particle turns. It can be shown that the instantaneous radius of curvature of the trajectory is proportional to the inverse value of the spin \cite{flocking_and_turning}.

 Equations~\eqref{eq:ism_plan1} and \eqref{eq:ism_plan2} can be rewritten in terms of the phase only, giving
 \begin{equation}
\chi \ddot{\varphi}_i+\eta \dot{\varphi}_i = 
  - \pdv{\cor{U}}{{\varphi}_i}+ \xi_i \, .
\label{eq:ism-2order}
\end{equation}
In the overdamped limit $\chi/\eta\to 0$ this second order equation reduces to a first-order one, which coincides with the planar Vicsek continuous model derived from \cref{eq:vicsek_v}. The ISM is therefore an inertial generalisation of the Vicsek model, which reduces to the latter in the dissipative limit.
The opposite limit in which $\eta\to 0$ renders a deterministic equation with a Hamiltonian reversible structure (see \cref{eq:ism_plan1,eq:ism_plan2}). In this case, since $s_i$ represents the generator of the rotational symmetry of the velocity, its total value $S=(1/N)\sum_i s_i$ is a conserved quantity.

\subsection{Collective turns and dynamical regimes}
\label{par:dynamical_regimes}
Let us now discuss how the ISM can appropriately describe the dispersion law observed in flocks of birds \cite{information_transfer}. To do so, we recall first what is known from experiments \cite{information_transfer,attanasi2015emergence}:
\begin{itemize}
    \item[-] bird flocks are highly ordered systems, with measured polarization values $\Psi \sim 0.9$;
    \item[-] turns are planar, \ie{} each trajectory lies approximately on a $2d$ plane;
    \item[-] mutual distances do not change during turns, and individuals sweep equal radius paths;
    \item[-] turns have a localized spatial origin, and the signal propagates linearly in time through the group with speed $c_s \gg v_0$ and with negligible attenuation.
\end{itemize}
Given these premises, we can now focus on the planar version of the model, as in \cref{eq:ism-2order}. This expression becomes even simpler if the system is in the deeply ordered phase, as flocks are. In this case, if we choose as a reference direction the one of the global velocity $\vb{V}$, the individual phases are very small, \ie{} $\varphi_i\ll 1$. 
One can then expand the potential $\cor{U}$ in \cref{eq:heisenberg} up to quadratic order -- which is called the {\it spin-wave} approximation -- to find
\begin{equation}
\cor{U}=- \sum_{ij} n_{ij} \cos(\varphi_j -\varphi_i)  \sim \sum_{ij} \Lambda_{ij} \varphi_j  \varphi_i  \ ,
\end{equation}
where we have introduced the discrete Laplacian $\Lambda$ with
\begin{equation}
    \Lambda_{ij} = -n_{ij} + \delta_{ij} \sum_k n_{ik} \, .
\label{eq:discretelaplacian}
\end{equation}
Since individuals do not change their mutual positions during turns, we can assume that the interaction network $n_{ij}$ remains approximately the same. If the phases vary slowly from one individual to the next, we can further treat phases as a continuous field $\varphi_i\to \varphi(\vb{x})$. The discrete Laplacian then becomes a continuous one, $\Lambda_{ij}\to -a^2 \grad^2$ ($a$ being the mean nearest-neighbour distance), and \cref{eq:ism-2order} becomes
\begin{equation}
    \chi \ddot{\varphi} + \eta \dot{ \varphi} -a^2 J n_c \nabla^2 \varphi=\xi \, ,
    \label{eq:phi_ddot}
\end{equation}
with $n_c=\sum_j n_{ij}$ (assumed to be constant on a regular network). 
While computations can be easily performed also in terms of the discrete variables, the continuous form is more convenient for visualizing the dispersion relation. Indeed, let
us now set $\xi=0$ into \cref{eq:phi_ddot} (or equivalently focus on $\expval*{\varphi}$), and Fourier transform in both space and time to get
\begin{equation}
    \chi \omega^2 -i\eta \omega - a^2 J n_c k^2=0 \, .
\end{equation}
From this relation we can immediately see that in the Vicsek case ($\chi=0$) the dispersion law becomes purely diffusive, \ie{} $\omega \sim i k^2$, leading to strong attenuation and a non-linear relation between space and time. Conversely, when $\chi$ is different from zero a more complex dispersion law is obtained,
\begin{equation}
    \omega_{\pm} = i\gamma \pm c_s \sqrt{k^2 - k_0^2} \, ,
    \label{eq:dispersion_relation}
\end{equation}
where we introduced the propagation speed $c_s$ and the effective dissipation $\gamma$ as
\begin{align}
    c_s &= a \sqrt{Jn_c/\chi} \, ,   \label{eq:sound_speed}\\
     \gamma&=\frac{\eta}{2\chi} \label{eq:gamma} \ ,
\end{align} 
and where $k_0\equiv \gamma/c_s$.

From \cref{eq:dispersion_relation} we can clearly evince that the two conditions observed in flocks of birds, \ie{} linear dispersion law and no attenuation, are reproduced when $\gamma \ll 1$ and $k_0\ll k$. In this deeply underdamped regime, the ISM predicts a linear dispersion law with propagation speed 
given by \cref{eq:sound_speed}. This is a highly non-trivial relation linking together the way directional information propagates through the system, and the degree of order (set by $J$). Remarkably, this prediction is very nicely verified in experimental data \cite{information_transfer}, thus supporting the idea that the ISM in the strongly underdamped regime provides a good description for collective turning in flocks.

More generally, according to the ratio of inertia and dissipation, and depending on the size of the system, the ISM interpolates between the dissipative Vicsek limit, and the efficient limit of flocks.  For $k \ll k_0$ all the modes are overdamped, while for $k \gg k_0$ the system sustains linear propagation with speed $c_s$ and constant attenuation with damping time $\tau = \gamma^{-1}$. Since  $k_0$ is the inverse of a length scale, it sets a limit on the size $L$ of a flock through which a signal can linearly propagate: indeed, imposing $k_{min}=L^{-1} \gg k_0$ implies $L \ll c_s/\gamma$. 
Another equivalent way to understand this constraint comes by thinking in terms of timescales. Indeed, the time needed for a signal of speed $c_s$ to cross the entire flock of size $L$ is
\begin{equation}
    \tau_s = \frac{L}{c_s}=\frac{L}{a}\sqrt{\frac{\chi}{J n_c}} \, ,
    \label{eq:tau_s}
\end{equation} 
but the same signal gets attenuated over a time scale 
\begin{equation}
\tau=\gamma^{-1} \ .
\end{equation}
We can then identify two dynamical propagation regimes:
\begin{itemize}
\item \textit{Underdamped regime (inertial propagation)}: $\tau_s < \tau$. In this case directional changes travel linearly though the whole system before attenuation can deteriorate the signal. Flocks of birds observed in experiments \cite{information_transfer} belong to the deep extreme of this regime, where attenuation is negligible for all the individuals. In terms of the parameters of the model, the condition defining this regime is $L^2 \eta^2/(4 a^2 J n_c \chi) < 1$.
\item \textit{Overdamped regime (dissipative propagation)}: $\tau_s>\tau$.  Propagation of information is inefficient because the signal gets dissipated before reaching the other end of the group. In this case, the signal travels unaffected up to certain scales $k^{-1}<L$, but it is damped on larger scales. This might result in a strong deformation of the group, or even in its fragmentation. It is likely that fish schools investigated in \cite{Rosenthal_2015} belong to this regime. 
\end{itemize} 
The dispersion law and the way information is propagated through the system determine the dynamical behaviour of the scalar polarization and of the correlation functions \cite{Review_2018}. In this work, we are interested in understanding how collective turns might arise in finite groups, and we shall find that the dispersion law affects in a crucial way the occurrence of these events. To do so, we will study the response of finite systems to a local perturbation under different boundary and motility conditions, and we will explore the dynamical regimes numerically by tuning the various parameters of the model.

\section{Perturbation events}
\label{par:eq_onlattice}
In the previous Section we introduced the ISM and we discussed the predictions of the model, under specific approximations, concerning the dispersion law in the system. In this Section, we want to generalise our discussion to the study of perturbation events. When a group changes its global direction of motion, this often happens because some external perturbation or disturbance acts upon one or more individuals in the group. This is not the only possibility (see \eg{} \cite{Cavagna_2017}), but it is certainly one of the most interesting. 
A reasonable choice is to model such a perturbation as a field $\vb{h}_i(t)$ linearly coupled to the individual velocities, in which case
\begin{equation}
    \cor{U} \to  \cor{U}_h=  \cor{U} - \frac{1}{v_0}\sum_{i} \vb{h}_i \cdot \vb{v}_i \, .
    \label{eq:coupling_hamiltonian}
\end{equation}
We note that, in principle, in the ISM it would also be possible to couple a field to the spins $\vb{s}_i$. This would however change the angular velocity of the perturbed individuals, while it would not bias them directionally, which is here our main purpose (see however \cite{tesi_venturelli}). 
In the planar case, the equations of motion then read
\begin{align}
&\dot{\mathbf r} = {\mathbf v}_i 
\ , \label{eq:ism_planr}\\ 
    &\dot{\varphi}_i = \frac{s_i}{\chi} \, , \label{eq:ism_plan1-2}\\
    &\dot{s}_i = 
      - \pdv{\cor{U}_h}{{\varphi}_i}
    -\dfrac{\eta}{\chi} s_i + \xi_i \, .
    \label{eq:ism_plan2-2}
\end{align}
These are the equations that we will consider in the numerical simulations performed in the remaining of this work. To make analytical progress, however, we still need to perform some simplifying approximations. 

\subsection{High order and slow network}
First of all, we consider a system in its polarised phase. Indeed, our aim is to investigate how a state of collective motion can change under external perturbations. We then perform the following approximations:
\begin{itemize}
    \item \textit{Spin-Wave Approximation (SWA)}: the system is highly polarized, and the individual velocity vectors $\vb{v}_i$ deviate weakly from the orientation of the global polarization $\vb{V}$. If the latter is chosen to be initially aligned to the $x$-axis when the external field is applied, then $\varphi_i\ll 1$ in the initial phases of the collective turn.
    \item \textit{Fixed network}: we assume that the adjacency matrix $n_{ij}$ appearing in the equations of motion is no longer a function of time. This assumption is reasonable when the timescale over which the collective turn develops is much shorter than the typical reshuffling time of the interaction network $n_{ij}$.  This condition of {\it local equilibrium} \cite{Nature_2016} is precisely what happens in flocks of starlings. A deviation from this condition is expected to arise for sufficiently large values of $v_0$. This analysis will be the subject of \cref{par:offlattice}.
\end{itemize}

Under these approximations, the equations of motion for $s_i$ and $\varphi_i$ take the form
\begin{align}
    \dot{\varphi}_i &= \frac{s_i}{\chi} \, ,
    \label{eq:eqn_mot_SWA_1} \\
    \dot{s}_i &\simeq - J \sum_{j} \Lambda_{ij} \varphi_j -\varphi_i h_i \cos \alpha_i + h_i\sin \alpha_i -\dfrac{\eta}{\chi} s_i + \xi_i \, , \label{eq:eqn_mot_SWA_2}
\end{align}
where $\alpha_i$ denotes the direction of $\vb{h}_i$, while $h_i=h_i(t)$ is its magnitude. We choose as the relevant observable the average polarization angle
 \begin{equation}
    \Phi(t) = \frac{1}{N} \sum_{i} \expval{ \varphi_{i} (t)} \, .
    \label{eq:polarization_angle}
\end{equation}
This $\Phi(t)$ coincides, for small angles $\varphi_{i}$, with the angle formed by the polarization measured at time $t$ after the perturbation, with the initial polarization (\ie{} the one maintained in the absence of perturbations). 
It therefore characterises the way the system changes in time its collective flight direction.

\subsection{Analytical results}
\label{par:perturbation_result}
The derivation of $\Phi(t)$ within the microscopic approach presented above is carried out in detail in \cref{app:derivation_microscopic} for a generic adjacency network $n_{ij}$; an alternative derivation, for a regular lattice and in terms of coarse-grained fields $\varphi(\vb{x},t)$, is presented in \cref{app:derivation_coarsegrained}, leading to the same result. Here we simply sketch the main steps, and comment the final expression, which will be compared to numerical simulations in \cref{par:onlattice}.

Equations~\eqref{eq:eqn_mot_SWA_1} and \eqref{eq:eqn_mot_SWA_2} represent a set of coupled linear first-order equations. They can be solved by  representing the phases $\varphi_i$ in the basis of eigenfunctions of the discrete Laplacian (see \cref{app:derivation_microscopic}). The solution reads
\begin{align}
    \varphi^i (t) = \sum_{a,j=1}^N \frac{ U\indices{^i_a} (U^{-1})\indices{^a_j}}{\chi\omega_a} &\int_0^t  \dd{t'} e^{-\gamma (t-t')} \sin[\omega_a(t-t')] \n\\ 
    &\times \left[\xi_j(t') + h_j(t') \sin \alpha \right]  ,
    \label{eq:main_intermediate}
\end{align}
where $\omega_a = \sqrt{J\lambda_a/\chi-\gamma^2}+\order{h}$,
 the columns of the matrix $U$ are the normalized eigenvectors of the discrete Laplacian introduced in \cref{eq:discretelaplacian}, and $\lambda_a$ are its eigenvalues.
Exploiting the properties of the eigenvectors (stemming from translational invariance), we get
\begin{equation}
    \Phi (t) = \frac{1}{N}\sum_{j=1}^N \int_0^t \dd{t'} e^{-\gamma (t-t')} 
    \frac{ h_j(t')\sin\alpha_j}{\chi \omega_0} \sin \omega_0(t-t') \, ,
\label{eq:Phi_t_nonlocal_main}
\end{equation}
where $\omega_0 = \omega_{a=0}= i \gamma +\order{h}$. Equation~\eqref{eq:Phi_t_nonlocal_main} should be considered valid within linear response for small $\vb{h}$. However, its validity extends to higher orders by choosing $\alpha_j\equiv \pi/2$, as we show in \cref{app:derivation_response}. This is the choice we will adopt in our numerical simulations in the following Sections.

Let us now specialize \cref{eq:Phi_t_nonlocal_main} to the case that interests us most. As we observed in \cref{par:vicsek}, collective turns in real bird flocks generally present a well-localized origin, as if in response to some punctual external stimulus, and only then they propagate throughout the group. This indicates that the perturbing field should be chosen as \textit{local}, \ie{} $\vb{h}_i\propto \delta_{ip}$, where $p$ labels a specific particle chosen inside the flock.
Applying a local field $\vb{h}_p$ on the particle labeled by $p$ induces on the mean polarization defined in \cref{eq:polarization_angle} a response
\begin{equation}
    \Phi(t) = \frac{1}{N\eta} \int_{0}^{t} \dd{t'} h_p(t') \sin \alpha_p(t') \left[ 1 - e^{-2 \gamma (t-t')} \right] \, .
    \label{eq:Phi_general}
\end{equation}
In the following, we will choose a step-like perturbation in the form $h_p (t)=A_0 \Theta(t)$ applied at a fixed angle $\alpha_p(t)\equiv \alpha_p$, for which \cref{eq:Phi_general} reduces to
\begin{equation}
    \Phi(t) = \frac{A_0 \sin \alpha_p}{\eta N}  \left[ t - \dfrac{ 1 - e^{-2 \gamma t} }{2 \gamma } \right] \, .
    \label{eq:response}
\end{equation}
Equation~\eqref{eq:response} tells us that, when the size of the system becomes very large (\ie{} in the thermodynamic limit), the system will never change the direction of the global order parameter on observational time scales. This is what happens in a ferromagnet at equilibrium: the order is stable at low temperature, and a local field applied on one site cannot change the total magnetization. However, biological groups have much smaller sizes than a physical condensed matter system. In this case, times $t\sim \mathcal{O}(N)$ can in fact be small enough to be reachable in experiments, and the system can thus change its global direction on observational time scales. 
We also note that, according to \cref{eq:response}, the group changes its direction independently of whether the propagation of information is inertial or dissipative. This last feature is however specific to the fixed-network condition: in this case the perturbed individual does not move, and it is always interacting with the rest of group. Even though information arrives damped as it travels through the system, it is injected continuously at the perturbed site and sooner or later everyone will turn. As we will discuss later on, this is not what happens for finite groups of moving individuals, when -- if information does not propagate unaltered and quickly enough -- the perturbed individual will leave the group before everyone can follow (see \cref{par:openBC}). 

The shape of the turn is described by \cref{eq:response}: for times $t\ll \gamma^{-1}$ the polarization angle $\Phi (t)$ grows with a quadratic dependence, while at larger times a linear behaviour is predicted. However, we remind that \cref{eq:response} is valid only for small phases $\varphi_i$, and it describes the polarization angle at most up to the time at which the turn is complete (\ie{} the angles clearly do not increase indefinitely, but a saturation occurs, which is not captured by our equation). Therefore, it is convenient to introduce the timescale of the collective turn $\tau_\T{turn}$, defined as the time at which $\Phi(t)$ becomes of $\order{1}$,
and which
represents the observational time window of interest for our analysis. According to the value of  $\tau_\T{turn}$ as compared to the typical timescales of the model, one might actually observe only linear or quadratic behaviour before the saturation:
\begin{itemize}
\item \textit{Linear behaviour}. If $\tau_\T{turn}\gg \gamma^{-1}$, then we can disregard the exponential decay in \cref{eq:response} to find
\begin{align}
    \Phi(t) &\simeq \frac{A_0 \sin \alpha_p}{\eta N} t \, ,    \label{eq:Phi_pred_over}\\
    \tau_\T{turn} &\simeq \frac{\eta N}{A_0 \sin{\alpha_p}} \ . \label{eq:tau_turn_over}
\end{align}
This condition is met for $\eta^2 N /(2\chi A_0 \sin{\alpha_p})\gg 1$, and it is typically what happens when dissipation is large with respect to inertia.
\item \textit{Quadratic behaviour}. The opposite regime is found when, on the contrary, $\tau_\T{turn}\ll \gamma^{-1}$. In this case, we can expand the exponential in \cref{eq:response} for small arguments and we find
\begin{align}
    \Phi(t) &\simeq \frac{A_0 \sin \alpha_p}{2 \chi N} t^2 \ ,     \label{eq:Phi_pred_under} \\
     \tau_\T{turn} &\simeq \left ( \frac{\chi N}{A_0 \sin\alpha_p}\right )^{1/2} \, .
     \label{eq:tau_turn_under}
\end{align}
This conditions is met for $\eta^2 N /(2\chi A_0 \sin{\alpha_p})\ll 1$, and it is typically what happens for very underdamped systems, in which inertia dominates over dissipation. In this regime, though, when the angle $\alpha_p$ is different from $\pi/2$, the expression for $\Phi(t)$ is slightly more complex (see  \cref{app:derivation_microscopic}). 
\end{itemize}
Finally, let us note that the conditions for the turn to appear linear or quadratic in the phase growth $\Phi(t)$ (which are stated above, and which involve the field amplitude $A_0$) are not exactly the same as the conditions for underdamped/overdamped propagation reported in \cref{par:dynamical_regimes} (which involve instead the interaction strength $J$). One can thus envision realizations of the system in which propagation is underdamped, but the turn still appears linear. As we mentioned before, the linear/quadratic behaviours represent the two extreme limits of \cref{eq:response}, which describes the polarization response in general; we distinguished these two behaviours mainly for practical reasons, since they are useful to check numerically the theoretical predictions, as we will do in the next sections.

\subsection{Other relevant timescales}
The behaviour of the polarization angle described by \cref{eq:response} is the immediate consequence of a very general feature of the potential $\cor{U}(\{\vb{v}_i\})$, namely the rotational invariance of the velocity-velocity interactions. In the low-noise phase, the system polarizes in a well-defined direction, thus breaking the symmetry. 
As in standard O($n$) models, the system remains however subject to soft Goldstone modes, \ie{} low-energy excitations, in the subspace perpendicular to the polarization vector. In other terms, due to the presence of noise, the vectorial polarization freely fluctuates like a random walk within this zero-mode subspace (see \cref{app:wandering}). 
When an external field is applied, it provides a bias to this random walk, which results in the (almost) linear time dependence exhibited by \cref{eq:response}. 

However, the spontaneous fluctuations provide another reference time scale. In fact, even in the absence of a field, a finite system subject to fluctuations will eventually depart from its original direction if we wait for long enough. In \cref{app:wandering} we compute explicitly this {\it wandering} time $\tau_w$, which grows with the system size $N$ and is regulated by the amplitude of the noise $T$. 
For times $t>\tau_w$, fluctuations have generally already changed (randomly) the direction of the polarization, and it becomes meaningless to speak of perturbation-response events. A condition to be satisfied in our analysis is therefore $\tau_\T{turn}\ll\tau_w$. 
Since the 
polarization response depends on the field amplitude, for this to happen we roughly need $A_0\gg T$ (see \cref{eq:tau_turn_over,eq:tau_turn_under}), a requirement which is easily achieved for very ordered systems. More accurate estimates of this condition are provided in \cref{app:wandering}.

If the system lives in two dimensions (\ie{} $d_v=2 \  \& \ d=2$), when the network is kept fixed the model is analogous to the XY model. This is not the case for flocks (which perform planar turns but are really three-dimensional systems), but 
we will adopt this simplification in
the numerical simulations discussed in the next sections.
It is known that the XY model does not exhibit long-range order in the thermodynamic limit. However, for finite-sized systems the order parameter (the magnetization, equivalent to the polarization defined here) remains finite in the low-temperature phase, and it slowly decreases for $N\to\infty$. In particular, for sizes comparable to those considered here, the system is fully ordered at low temperature (see \cref{app:mermin_wagner}).

\begin{figure*}[t]
\centering
  \includegraphics[width=\textwidth]{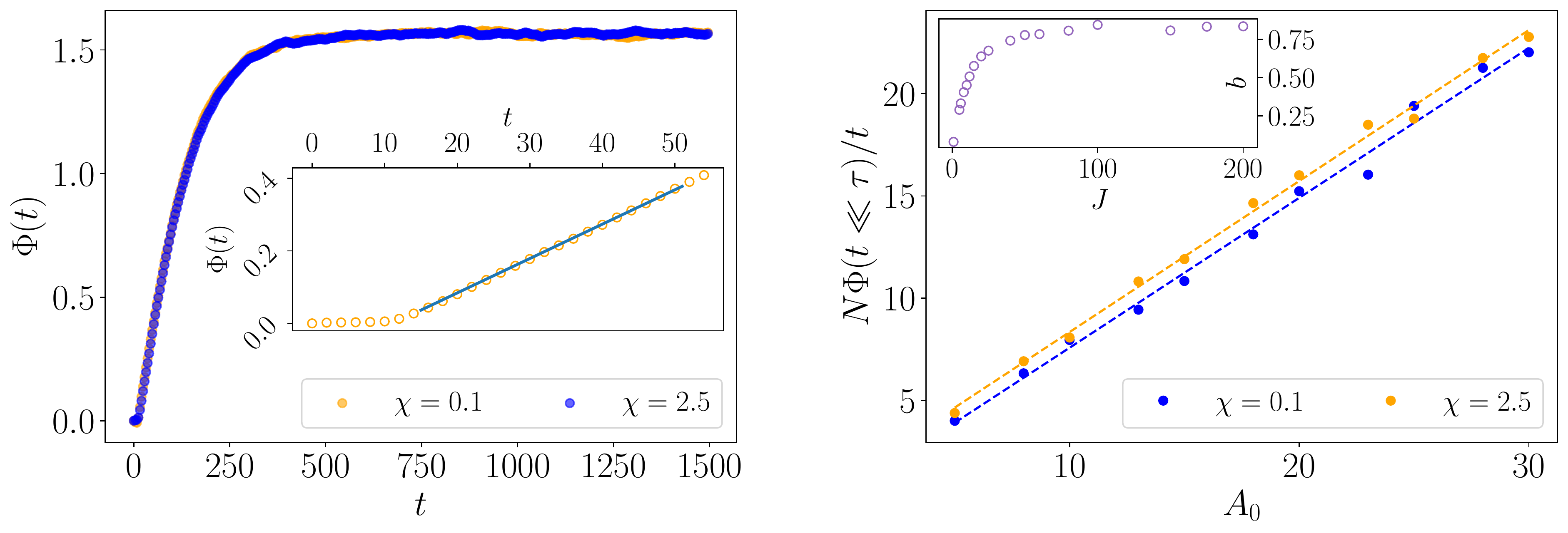}
    \put(-505,162){a)} 
    \put(-235,162){b)}
    \\
  \includegraphics[width=\textwidth]{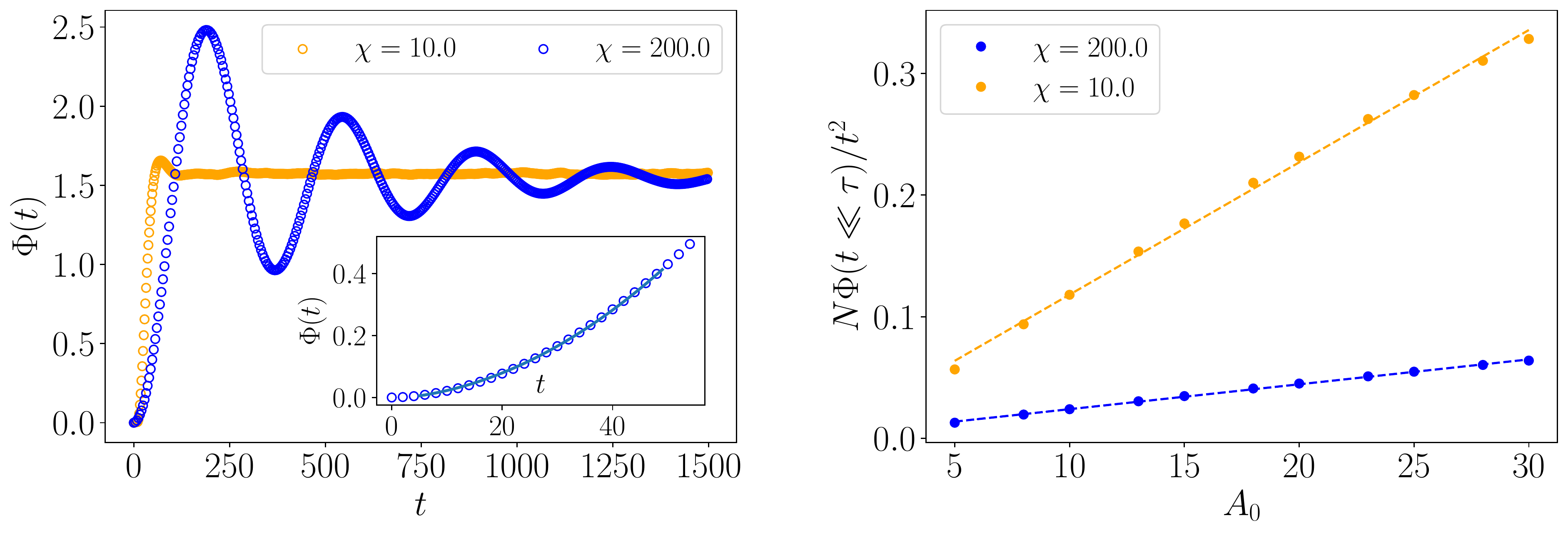}
  \put(-505,162){c)}
  \put(-235,162){d)}
\caption{Linear behaviour (panels (a) and (b)) and quadratic behaviour (panels (c) and (d)) during collective turns, in the on-lattice model. \textbf{(a)} Mean polarization angle $\Phi(t)$
in the regime $\tau_\T{turn}> \gamma^{-1}$. After a short transient (see text), the dependence of $\Phi$ on time $t$ becomes linear, before reaching saturation along the direction singled out by the external field (here $\alpha_p =\pi /2$). 
The two values $\chi =0.1$ or $\chi=2.5$ correspond to $\tau_\T{turn}\gg \gamma^{-1}$  or $\tau_\T{turn}\sim \gamma^{-1}$, respectively; we used $A_0=5$.
Inset: zoom on the small-$t$ region of the curve, which we fit using a straight line.
 \textbf{(b)} Slope of the polarization angle (multiplied by $N$) as a function of the field amplitude $A_0$. For each value of $A_0$, the slope is extracted from the linear fit of the polarization angle curve (see inset of panel (a)). The linear dependence predicted by \cref{eq:Phi_pred_over} is very well obeyed. The slope of the curves is slightly different from  the predicted one, due to the spin-wave approximation; however, the slope actually approaches $\sin \alpha_p /\eta=1$ (with our choice of parameters) as the interaction strength $J$ increases, whence the SWA becomes more reliable (as we show in the inset).
 \textbf{(c)} Mean polarization angle $\Phi(t)$ 
 in the regime $\tau_\T{turn} < \gamma^{-1}$. After an initial quadratic growth, $\Phi (t)$ saturates to the asymptotic value $\pi/2$. In the deeply underdamped regime, $\Phi (t)$ exhibits damped oscillations before coming to rest along the direction of the external field. 
 The two values $\chi=10$ and $\chi=200$ correspond to $\tau_\T{turn} \lesssim \gamma^{-1}$ or $\tau_\T{turn} \ll \gamma^{-1}$, respectively; we used $A_0=30$.
 \textbf{(d)} Quadratic growth coefficient of the polarization angle (multiplied by $N$) as a function of the field amplitude $A_0$. For each value of $A_0$, the coefficient is extracted from the quadratic fit of the small-$t$ region of the polarization angle curve (see inset of panel (c)). 
 The linear dependence predicted by \cref{eq:Phi_pred_under} is very well obeyed. 
 In all the plots we used $T=0.005, J=50, \eta =1, N=400$, and $\Phi(t)$ is averaged over $n_\T{turns}=10$ realizations of a turning event.}
    \label{fig:collective_turns}
\end{figure*}

\section{Numerical results -- fixed network}
\label{par:onlattice}

In the previous Section we derived analytically, under some suitable assumptions, the response of the system to a local directional perturbation, \ie{} the time dependence of the polarization angle $\Phi(t)$ in \cref{eq:Phi_general}. In this Section we numerically test the validity of such predictions. 

We start by addressing the problem in the fixed-network case, where we can build up and validate a suitable perturbation protocol, which will be carried over to the off-lattice case in \cref{par:offlattice}. For simplicity, we consider as fixed network a regular two-dimensional lattice.
%
%

We implemented a time-discretized version of the ISM equations \eqref{eq:ism_planr}, \eqref{eq:ism_plan1-2} and \eqref{eq:ism_plan2-2} by generalizing the numerical integration scheme proposed in
\cite{ciccotti}, as reported in \cref{app:numerical_integration}. We adopted a step-like time dependence for the perturbation field $h_p(t)$ (see \cref{eq:Phi_general}), in the form 
\begin{equation}
    h_p(t)=\dfrac{A_0}{2}\Bigg[ 1 + \tanh\Big( \dfrac{t-t_0}{\tau_\T{step}} \Big)\Bigg] \, ,
    \label{eq:simil_step}
\end{equation}
where $A_0$ controls the amplitude of the field, $\tau_\T{step}$ controls the sharpness of the step, and $t_0$ is a time offset. As we described in \cref{par:eq_onlattice}, in our numerical simulations we apply the perturbing field on a single individual labelled by $p$ in order to mimic the spatially localized origin of turns in real flocks. We set for simplicity the angle between the individual and the field to be $\alpha_p = \pi /2$, so that the predictions in \cref{eq:Phi_pred_over,eq:Phi_pred_under} further simplify since $\sin \alpha_p=1$. We adopt for the moment periodic boundary conditions, to pinpoint the role of the dynamics. In Sec.~\ref{par:openBC} we will investigate the effect of open boundary conditions, which turns out to be crucial when the system is off-lattice.

The choice of the time scales $t_0$ and $\tau_\T{step}$ in \cref{eq:simil_step} plays no crucial role, as long as the field change remains sharp: we set $\tau_\T{step} = 0.1$ to approximate a step function, and $t_0 =10$ so that the system starts turning after a short transient of $\mathcal{O}(t_0)$ (but other choices would return similar results provided that $\tau_\T{step}$ is chosen not too large). Conversely, the range of values over which $A_0$ can be chosen requires some discussion. First, \cref{eq:Phi_pred_over,eq:Phi_pred_under} tell us that changing the amplitude $A_0$ of the perturbing field offers a way to tune the velocity of the turn: we thus avoid choosing too large values of $A_0$, as they would induce a turn which is too fast to be easily studied. 
Besides, large values of $A_0$ might push us away from the linear response regime that we wish to explore.
Conversely, a lower bound on $A_0$ is actually imposed by the physics of the system. Indeed, due to its finite size, the polarization angle $\Phi(t)$ is subject both to the action of the perturbing field, and to the wandering effects described in the previous Section (which are non-negligible even if we are working with a highly polarized state). The value of $A_0$ should then be chosen sufficiently large so as to avoid that -- on our time scales of observation, and for our limited number of trials -- the effect of the perturbing field gets completely masked by wandering effects. 
This translates to requiring that the polarization angle $\Phi(t)$ is of $ \mathcal{O}(1)$ for times $t \leq \tau_w$ in all the dynamical regimes described above. With our set of parameters, this condition is met for $A_0 \sim \mathcal{O}(1)$, hence we simulate our turns in the range $A_0 \in [5,30]$.

\begin{figure*}[t]
    \centering
    \includegraphics[width=0.9\textwidth]{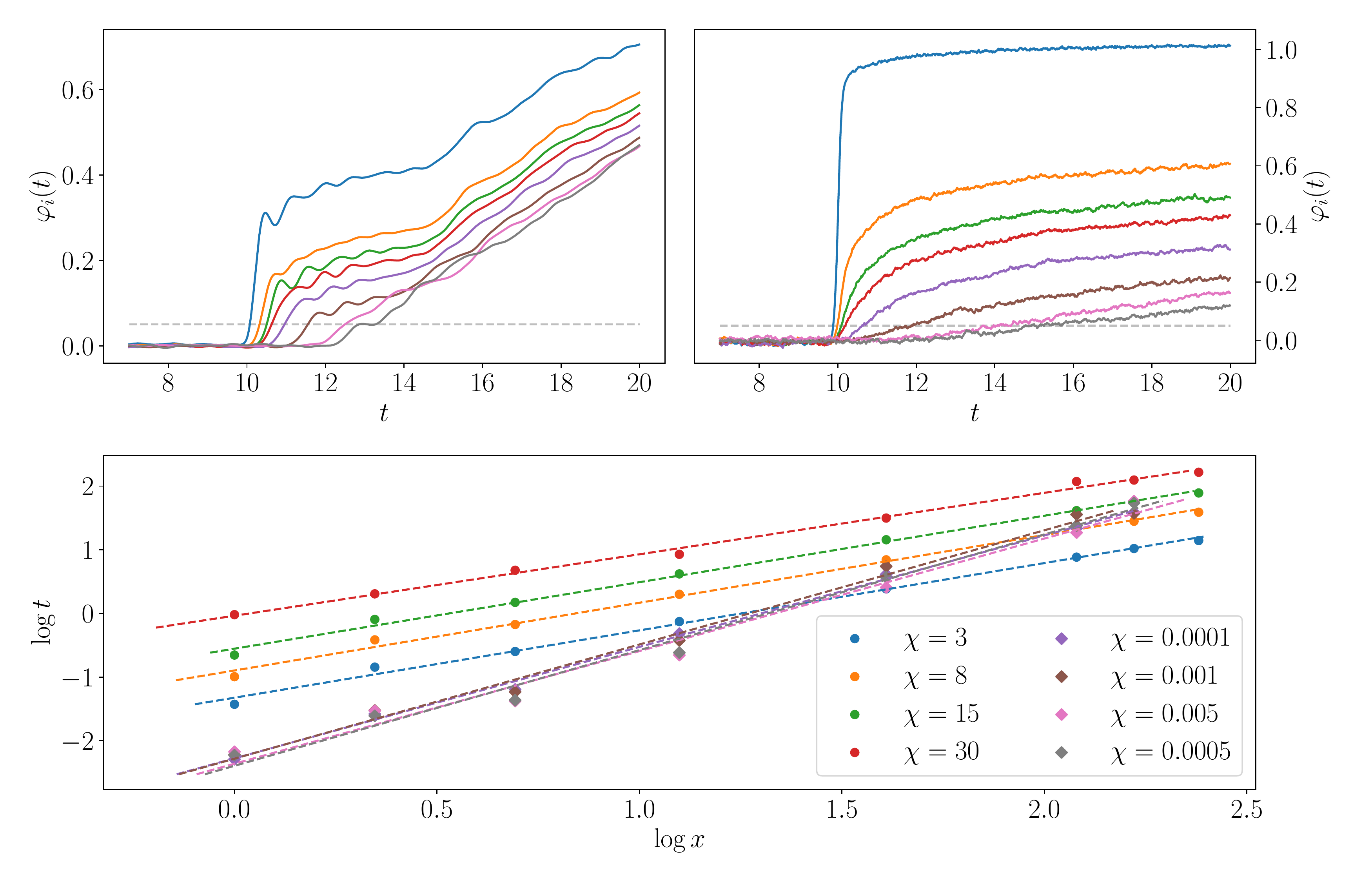}
    \put(-415,270){a)}
    \put(-220,270){b)}
    \put(-415,127){c)}
    \put(-355,105){$x\sim t$}
    \put(-270,55){$x\sim \sqrt{t}$}
    \caption{Information propagation and dispersion law. Panels \textbf{(a)} and \textbf{(b)}: individual phases as a function of time, for several individuals in the system. The first individual to turn (blue curve) is the one on which the local external field is applied. Other curves correspond to individuals located -- from left to right -- at $n=1,\sqrt{2}, 2, 3, 5, 8, \sqrt{85}, 9\sqrt{2}$ lattice spacings from the perturbed one. Panel (a) displays a perturbation event in the underdamped regime ($\chi \in [3,30]$), while panel (b) displays an event in the overdamped regime ($\chi \in [10^{-4},10^{-3}]$). All these curves eventually saturate to $\varphi_i=\pi/2$ at long times (not shown). Panel \textbf{(c)}: dispersion law for different values of the model's parameters. For each perturbation event, the position and turning time of each individual is extracted from the curves of the individual phases (as in panels (a) and (b), see the main text), and displayed in logarithmic scale. Each curve corresponds to a different perturbation event. We also used $T =0.005, \eta = 1, J=50, N=400$.
    }
    \label{fig:displaw}
\end{figure*}

\subsection{Collective turns}
\label{par:collective}
Here we describe our first simulations of collective turns. Hereafter, the parameters of the model are chosen so that the system is highly polarized ($\Psi > 0.98$, as in real starling flocks \cite{information_transfer}). This is useful for two reasons: first, it allows to obtain polarization curves that are not excessively noisy; second, by setting the flock in its deeply ordered state we are enforcing the conditions for the SWA, under which our predictions were derived (see \cref{par:perturbation_result}).

Let us introduce the following numerical protocol:
\begin{enumerate}[(i)]
\item the system of size $N$ is initialized on a square lattice of side $L= \sqrt{N}$ in a disordered configuration. We wait for the system to thermalize and then we measure the direction of the order parameter, which is chosen as the reference value, \ie{} $\Phi(t=0)\equiv 0$;
\item we select a value for the field amplitude $A_0$;
\item we apply the external field on a single individual $p$ inside the system, and we keep the field switched on for a time $T_\T{turn}$;
\item we switch off the field and let the system thermalize for a time $T_\T{term}$;
\item to obtain more robust observations, we repeat steps (iii) and (iv) for $n_\T{turns}$ times, and we average over the various realizations of the turning event.
\end{enumerate}

We can first use our numerical simulation to check the linear and the quadratic behaviour predicted for the polarization curves $\Phi(t)$, 
see \cref{eq:Phi_pred_over,eq:Phi_pred_under}. We keep the parameters $J$, $T$ and $\eta$ fixed, and we use the inertia $\chi$ as the tuning parameter to explore the different regimes of $\tau_{\rm turn}$ vs. $\gamma^{-1}$. 
Let us start with the case $\tau_{\rm turn}> \gamma^{-1}$, where our analytical derivation predicts a linear behaviour of the polarization angle $\Phi(t)$ -- see \cref{eq:Phi_pred_over}. A typical mean polarization curve obtained in this regime is shown in \cref{fig:collective_turns}a for two values of $\chi$, together with a linear fit of the initial part of the curve (see inset). Recall that, due to the SWA, the agreement with the analytical prediction is expected to break down as the polarization angle saturates to its final value $\alpha_p=\pi/2$. 
To test the agreement of the prefactor in \cref{eq:Phi_pred_over} with our numerical observations, we have repeated steps (i) to (iv) for several values of $A_0$ in the range $[5,30]$, and for each value of $A_0$ we have fitted the initial part of the mean curve $\Phi(t)$ with a linear function. Figure~\ref{fig:collective_turns}b shows our results for the same values of $\chi$ as in \cref{fig:collective_turns}a, which confirm the prediction of a linear dependence of the slope on $A_0$.

Keeping the values of the parameters $J$, $\eta$ and $T$ fixed, while pushing the inertia $\chi$ towards higher values, brings the system more into the underdamped region where $\tau_{\rm turn}<\gamma^{-1}$. Here the predicted time dependence for $\Phi(t)$ is quadratic, as described by \cref{eq:Phi_pred_under}. We note that an initial quadratic behaviour is actually expected to occur in any regime, since there is always a time window (however short) in which the condition $\gamma t \ll 1$ is met -- see \cref{par:perturbation_result}. However, as discussed in the previous section, the quadratic behaviour holds for the whole duration of the turn only when $\tau_\T{turn}\ll\gamma^{-1}$. This occurs if $\eta^2 N /(2\chi A_0 \sin{\alpha_p})\ll 1$, which translates to $\chi \gg 5$ for the parameters used in \cref{fig:collective_turns}c, where we show two polarization curves $\Phi(t)$ for two distinct values of $\chi$.

We then repeat the same analysis as for the linear case: for each value of $A_0$ we fit the initial part of the polarization curve $\Phi(t)$ with a parabola (see inset in \cref{fig:collective_turns}c), and check that the estimated prefactor grows linearly with $A_0$ (see \cref{eq:tau_turn_under}). This is precisely what happens, as displayed in \cref{fig:collective_turns}d.

Finally, a
curious feature of the deeply underdamped regime is the occurrence of oscillations in the saturation region of the polarization angle (see the blue curve in \cref{fig:collective_turns}c). This is intuitive once we recall that the parameter $\chi$ 
plays the role of a moment of inertia in \cref{eq:ism-2order}. A quantitative estimate (see \cref{app:oscillations}) predicts oscillations with frequency $\Omega=\sqrt{A_0/(\chi N)}$ and damping factor $\gamma$, in line with numerical observations. 

\subsection{Propagation law}
\label{par:propagationlaw}
Our numerical protocol additionally allows us to investigate and check the dispersion law predicted in \cref{par:dynamical_regimes}. The behaviour of the polarization angle $\Phi (t)$ discussed above describes how the group as a whole rearranges its direction to align with the external field. It does not describe, however, how the perturbation -- which is applied locally to a specific individual -- is propagated from individual to individual through the system. To explore this issue, we need to consider the phases $\varphi_i(t)$ of the single individuals, and monitor how they change in time. In Fig.~\ref{fig:displaw}a,b we thus show these individual curves as a function of time, for two different sets of the model's parameters. The perturbed individual (blue curve) is the first to change its direction and to complete the turn. Other individuals display a very similar turning profile, but shifted in time with a certain delay $\Delta t_i$, which is larger the farther the individual is from the location of the perturbation (colored curves from left to right). This behaviour indicates that the directional information spreads progressively through the system. Starting from these curves, we can quantify the dispersion law as follows. Let us identify the turning time of a given individual
as the time when its direction reaches a threshold angle
$\varphi=0.05$ (dashed horizontal line in Fig.~\ref{fig:displaw}a,b). 
The set of individuals which start turning at a given time $\Delta t_i$ determine the turning front at that time, and their 
distance $x_i$ from the perturbed individual 
identifies the location of the front, which travels obeying the dispersion law $x_i(\Delta t_i)$. 
We plot the latter in Fig.~\ref{fig:displaw}c for several simulations performed with different sets of parameters. The figure confirms the picture outlined in  Sec.~\ref{par:dynamical_regimes}, where two propagation regimes were identified according to whether the time for information to travel through the entire system, $\tau_s=L/c_s$, is smaller or larger than the time for the attenuation to damp the signal, $\tau=\gamma^{-1}$. When $\tau_s\ll \tau$ a linear dispersion law ($x\sim t$) is observed; conversely, when $\tau_s\gg \tau$ we find a dissipative behaviour ($x\sim \sqrt{t}$). With the parameters used in 
Fig.~\ref{fig:displaw}, one has $\tau/\tau_s=1$ for $\chi=0.5$. 


\begin{figure*}
\centering
\includegraphics[width=\columnwidth]{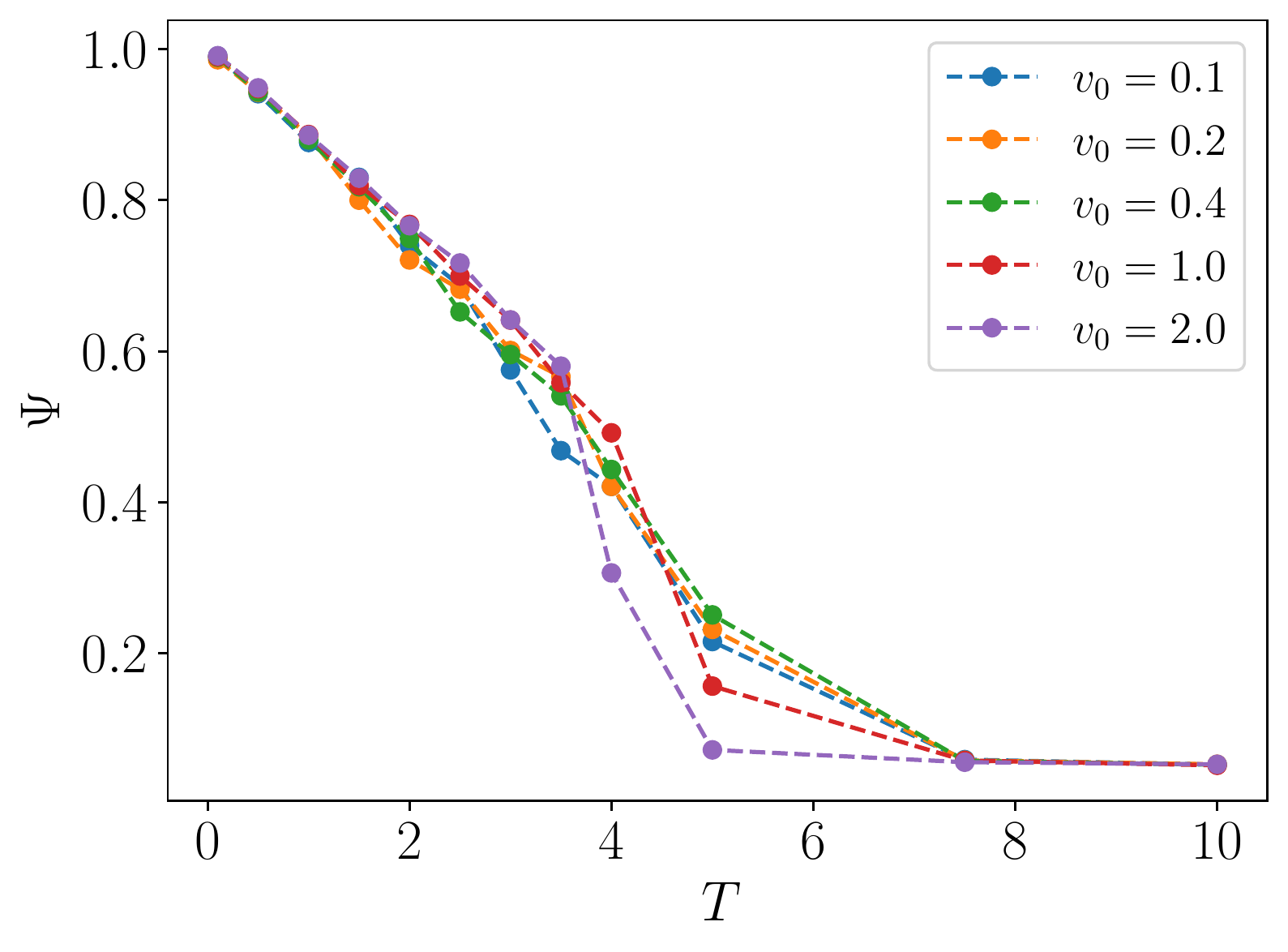}
\includegraphics[width=\columnwidth]{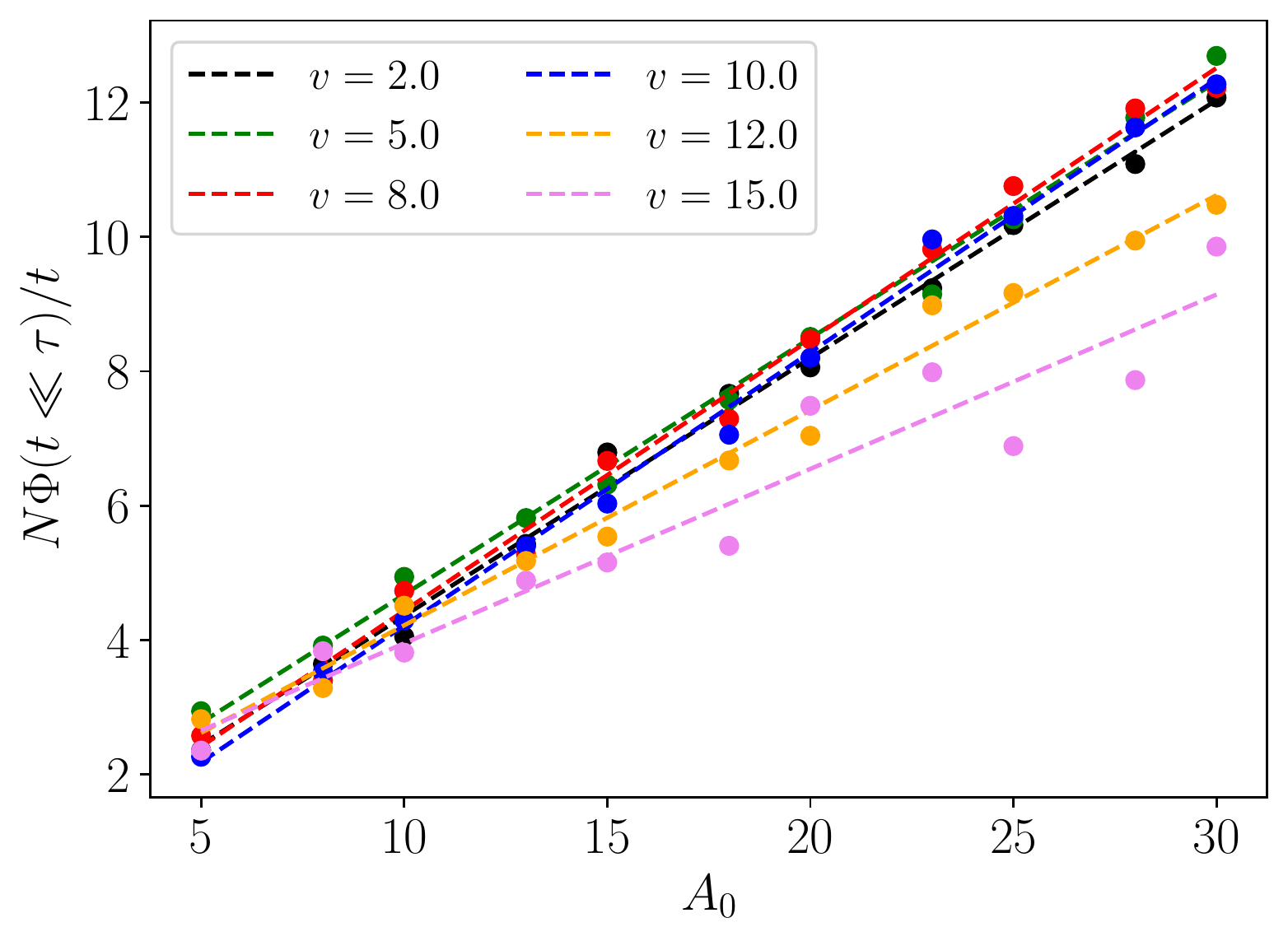}
\put(-85,32){\includegraphics[scale=0.045]{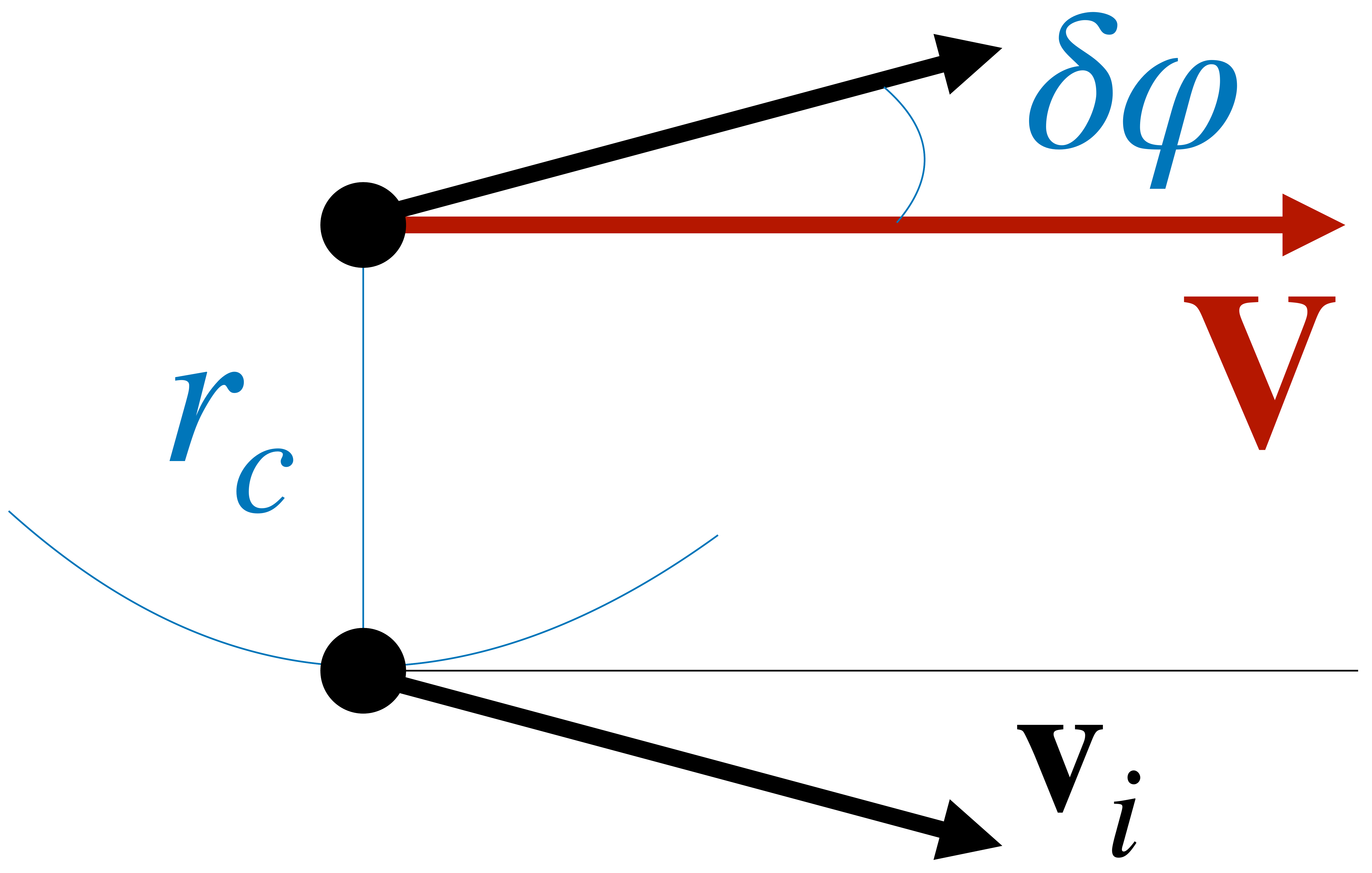}}
  \put(-495,170){a)}
  \put(-245,170){b)}
\caption{Simulations of the off-lattice model with periodic boundary conditions. Panel \textbf{(a)}: phase diagram of the model.
The scalar polarization $\Psi$ is measured as a function of the temperature $T$ for several values of the activity $v_0$, in the overdamped regime. The parameters used in the plot are $\chi=1.25, \eta = 5, J=0.8, N=400$. Panel \textbf{(b)}: effect of the perturbation.
As we did in \cref{fig:collective_turns} for the on-lattice case, we check the linear dependence of the polarization angle $\Phi(t)$ (predicted in the on-lattice case by \cref{eq:Phi_pred_over}), averaging over $n_\T{turns}=10$ realizations for each value of the activity $v_0$. The higher the activity, the more the fitted linear prefactor deviates from the prediction of \cref{eq:Phi_pred_over}, until for $v_0 = 15$ we are totally out the validity of the on-lattice derivation, because the network reshuffling is so fast that the fixed-network approximation no longer holds. We used the parameters $\chi = 5, J = 1, T = 0.01, N = 400,\eta=1$, with scalar polarization $\Psi \sim 0.98$. Inset: sketch showing two interacting individuals moving away from each other. Eventually their distance becomes larger than the interaction radius $r_c$. Here $\vb{V}$ indicates the common flight direction of the flock, while $\delta \varphi$ is the deviation of a single individual due to thermal fluctuations.} 
\label{fig:offlattice}
\end{figure*}

\section{Numerical results off-lattice}
\label{par:offlattice}

In this Section we address the question of how a local perturbation affects the global directional order in the full off-lattice model. 
In particular, we want to test whether and to which extent the analytical predictions derived in \cref{par:perturbation_result} are valid when the activity of the model is taken into account. Indeed, the presence of a nonzero activity $v_0$ introduces a new time scale in the model: its interplay with the previous timescales is expected to foster a rich phenomenology, as it happens already in Vicsek-like models \cite{chepizhko2021revisiting}. We will start by studying the unperturbed model, and then move on to analyze the effects of a local perturbation.

\subsection{Unperturbed model}
The numerical implementation of the off-lattice model is similar to that of the on-lattice case, but this time the individual positions $\vb{r}_i$ evolve ballistically -- according to the corresponding velocity vectors $\vb{v}_i$ -- in a squared box with periodic boundary conditions. The interaction network $n_{ij}(t)$ must now be updated regularly: this is obtained by implementing the \textit{cell-list} method (see \cite{frenkel,tuckerman,code} and \cref{app:numerical_integration}). We remind that we chose a metric interaction rule, \ie{} $n_{ij}$ is different from zero if individuals $i$ and $j$ have a mutual distance lower than the interaction range $r_c$. 

Studying the phase diagram of the model is made nontrivial by the presence of spatial aggregation effects. Even Vicsek-like models (\ie{} first-order models) are known to exhibit strong spatial heterogeneities, for sufficiently low noise, in the case of \textit{additive} interactions \cite{chepizhko2021revisiting}. By this we mean an interaction term (in the evolution equation for the individual velocity vector) of the form
\begin{equation}
    \sim \sum_{j \in \mathcal{N}_i} \sin ( \varphi_j - \varphi_i) \, ,
\end{equation}
which is not normalized by the total number $\cor{N}_i$ of particles interacting with the $i$-th (as it happens instead in non-additive models). As a result, the interaction becomes stronger and stronger as new particles enter the interaction range of the $i$-th, which might cause the formation of clusters. Since the equations of motion of our model feature an interaction term of this form (see \cref{par:perturbation_result}), we expect a similar mechanism to take place.

Clustering phenomena were indeed observed in our numerical simulations. We considered an initially uniform distribution of individuals; when a density fluctuation brings a few individuals close together, the interaction force increases and eventually dominates over the fluctuations, which are then unable to disrupt the cluster. A detailed analysis of the clustering formation goes beyond the scopes of this work, but one generally finds that these aggregation effects are more prominent at high 
activity, low temperatures (in the ordered phase of the system), and high values of the inertia (\ie{} towards the underdamped regime). On general grounds, we expect the timescale of the clustering process to be mostly influenced by the inertia $\chi$, which slows down the effective motility of the system.


The off-equilibrium phase diagram in \cref{fig:offlattice}a shows the scalar polarization $\Psi$ of the model as a function of the temperature $T$, for various speeds $v_0$. In constructing the phase diagram, we selected a range of dynamical parameters $(\eta,\chi)$ for which the most severe aggregation effects are absent.
As we noted, the density $\rho = N/L^2$ of the system and the interaction radius $r_c$ heavily affect clustering phenomena and, more in general, the dynamics of the model. Hereafter we work with $\rho=1$ and $r_c = 1.5$, which ensure (within non-clustered, homogeneous systems) an average number of interacting neighbours $n_c\sim \pi r_c^2 \rho \sim 7$. This value reproduces the one measured in real starling flocks, where each bird is found to coordinate
with its nearest $7-8$ individuals \cite{cavagna2015short}.

Note that, while numerical simulations of the $3d$ ISM have been previously reported in the literature \cite{flocking_and_turning,ferretti2020building,3.99}, the ones presented in this work are the first numerical simulations of the $2d$ ISM.


\subsection{Perturbed model: role of the dynamics}
\label{eq:offlattice_role_dynamics}
After analyzing the unperturbed case, we now address the role of the activity in the presence of a local perturbation. We noted in \cref{par:onlattice} that a turn \textit{always} takes place in a finite-size, on-lattice system in response to a step-like local perturbation. However, once we bring in a nonzero activity, we are no longer guaranteed that a local perturbation will produce a similar behaviour on the entire system. Indeed, two timescales are now expected to compete: that of information propagation (related to the speed $c_s$, see \cref{eq:sound_speed}), and the reshuffling time of the network $n_{ij}(t)$, which is the time taken by an individual to change its neighborhood. If the latter process is too fast, then a given individual may leave its interaction neighborhood before the \textit{signal} is able to propagate.  
We thus generically expect a threshold activity $v_0^\T{lim}$ such that, for $v_0<v_0^\T{lim}$, the off-lattice system behaves similarly to the on-lattice case, and our analytical predictions for the time dependence of the polarization angle $\Phi(t)$ still apply. 

We can estimate $v_0^{\text{lim}}$ by using a simple heuristic argument.
Consider two distinct interacting individuals, \ie{} at a relative distance smaller than the interaction range $r_c$. Let $\delta \varphi$ denote the angular deviation of their flight direction with respect to that of the flock $\vb{V}$, as in the inset of \cref{fig:offlattice}b. In the fully polarized limit $\Psi =1$ one has $\delta \varphi =0$, meaning that all the birds fly straight and never cross; conversely, in the presence of fluctuations $\delta \varphi$, one can derive under the SWA the relation (see \cref{app:fluctuations})
\begin{equation}
    \delta \varphi \equiv \sqrt{\expval{\varphi^2}} \sim \sqrt{2(1- \Psi)} \, .
    \label{eq:pol_fluc}
\end{equation}
In the worst-case scenario, the velocity vectors of the two individuals point outwards as in the inset of \cref{fig:offlattice}b, so that they drift away from each other and stop interacting after a time $t_r$ defined as 
\begin{equation}
2 v_0 \, \delta \varphi\, t_r \sim r_c \, .
\end{equation}
The timescale $t_r \sim r_c/(2 v_0 \, \delta \varphi)$ thus estimates the \textit{reshuffling} time of the connectivity matrix $n_{ij}(t)$.
When an external field acts on a single individual, the information must reach its nearest neighbours before reshuffling happens, in order for a collective turn to take place in the way it occurs on a fixed network. Such information propagates across the inter-particle distance $a$ within a time $t_a \sim a/c_s$, where $c_s$ is the information propagation speed introduced in \cref{eq:sound_speed}. We conclude that information propagates similarly to the on-lattice case if $t_a \ll t_r$, so that the network can be seen as quasi-fixed; in turn, this implies
\begin{equation}
    v_0 \ll \dfrac{r_c}{2a} \dfrac{c_s}{\delta \varphi} =  \dfrac{r_c}{2} \sqrt{\dfrac{J n_c}{2 \chi (1-\Psi) } }  \equiv v_0^{\text{lim}} \, .
    \label{eq:v_0_lim}
\end{equation}

\subsection{Numerical results}
Using the same protocol that we applied in the on-lattice case (see \cref{par:onlattice}), we now analyze the effects of a local perturbation. Again, the parameter $\chi$ can be tuned in order to explore the dynamical regimes of the model. The choice of the other parameters is instead dictated by two new requirements. First, we should make sure that the system is not in a clustered state.
We then check beforehand that the time scale of clustering is much longer than the turning time $\tau_\T{turn}$ of the flock; this requires us to tune the temperature $T$ in order to remain within a well-polarized state, where the system is spatially uniform. Secondly, numerical limitations (such as time discretization) would prevent us from exploring the region with $v_0 > v_0 ^{\text{lim}}$, if the latter were excessively high. All in all, 
this leads to the choice of parameters of \cref{fig:offlattice}b,
which corresponds to $\Psi \sim 0.98$ and $v_0 ^{\text{lim}} \sim 10$ according to \cref{eq:v_0_lim}.

We focus here as an example on the overdamped region $\tau_\T{turn}\gg \gamma^{-1}$, but similar results were observed also in the underdamped case. In this regime we know that the polarization angle $\Phi(t)$ should behave linearly (see \cref{eq:Phi_pred_over}): in \cref{fig:offlattice}b we thus fit its initial part with a straight line, and we plot the so-obtained slope against $A_0$ for several values of $v_0$. Most of these lines overlap, which is expected since \cref{eq:Phi_pred_over} is $v_0$-independent; conversely, the behaviour starts to deviate from a straight line as we approach the region with $v_0 \sim v_0 ^{\text{lim}}$. This marks the breakdown of the on-lattice approximation.


We finally note that, in the entirety of this Section, the activity $v_0$ of the system does not seem to affect the ability of the local field to trigger a collective reorientation, at least for speeds not much larger than $v_0 ^{\text{lim}}$ (we explored values up to $v_0=30$); indeed, the mean polarization angle always saturates to $\Phi (t \to \infty) \to \pi/2$, even in the overdamped regime where propagation of information is attenuated by damping.
The reason why this happens is crucially related to the presence of periodic boundary conditions: even if the individual feeling the perturbation moves, it never leaves the system, and the perturbation therefore keeps acting on the 
latter indefinitely.
This is not the case, on the contrary, when open boundary conditions are considered. In the next section we will investigate these conditions, where -- as we will discuss -- the impact of the activity and the regime of information propagation are crucial in determining the occurrence of a turn.

\begin{figure}[t]
    \centering
       \includegraphics[width=\columnwidth]{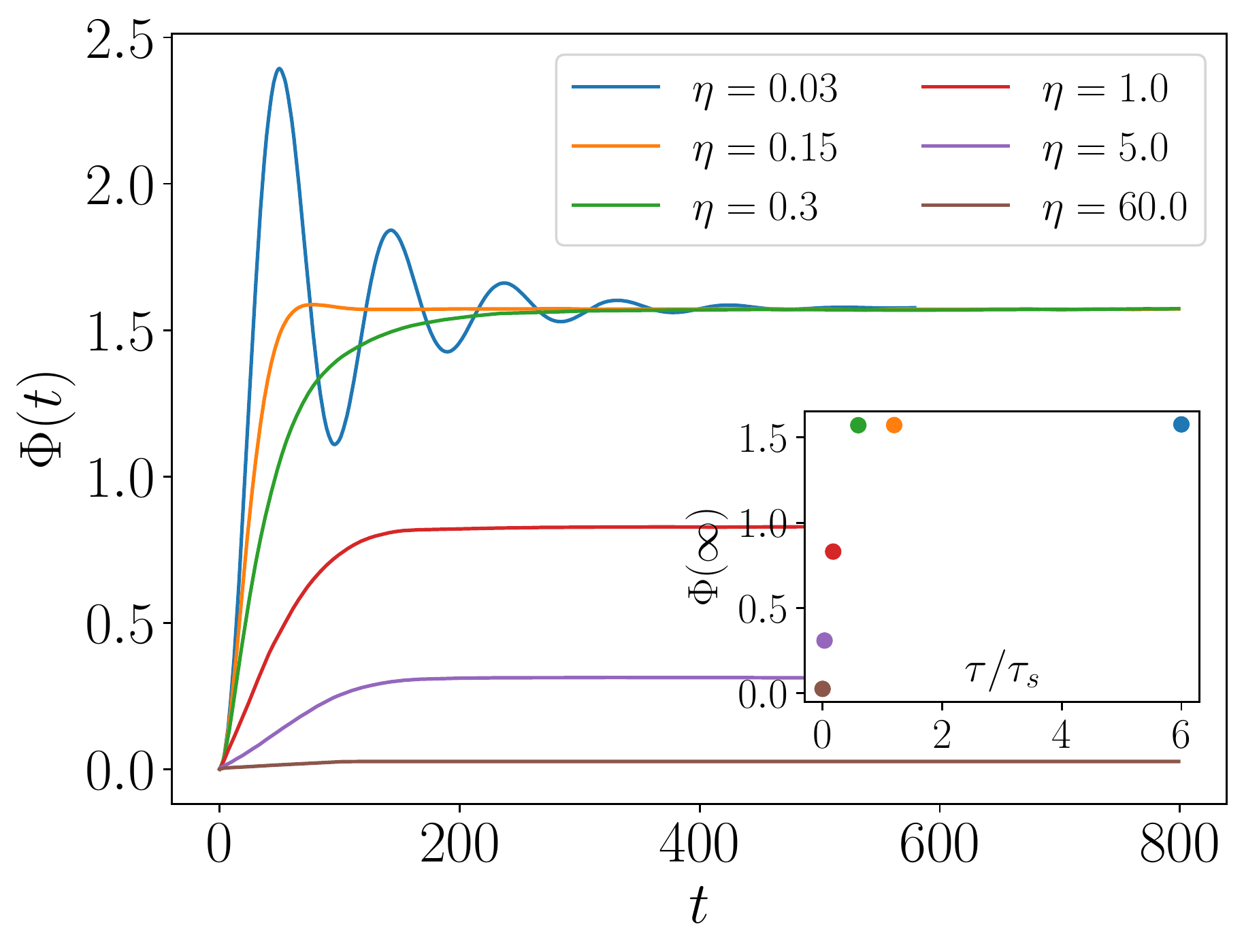}
    \caption{Mean polarization angle in off-lattice simulations with open boundary conditions, for several distinct values of the parameter $\eta$. Only the systems with $\eta \leq 0.3$ (corresponding to the underdamped propagation regime) are able to perform a complete collective turn. Inset: limiting polarization angle as a function of the ratio $\tau/\tau_s$. The plateau value tends to $\pi/2$ when passing from the overdamped to the underdamped regime. The parameters used in the plot are $\chi = 1.25, J = 0.8, T = 8 \times 10^{-5},  v_0 = 0.1, N = 400, A_0 = 30, n_\T{turns}=20$. 
    }
    \label{fig:OBC_phi_t}
\end{figure}

\begin{figure*}
    \centering
    \includegraphics[width=\textwidth]{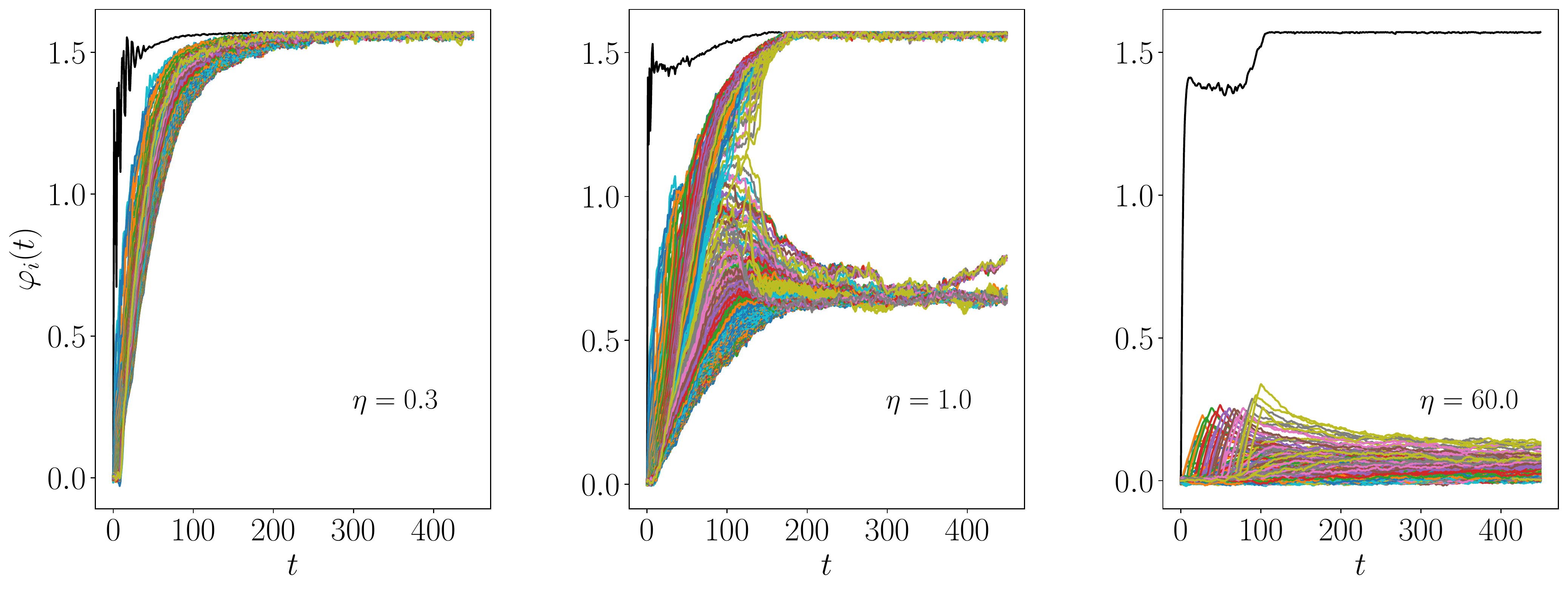}
    \caption{Evolution of the polarization angle $\varphi_i(t)$ of all the single components of the flock, along a single realization of a turning event performed with open boundary conditions (see \cref{sec:obc_regimes}). In all the figures, the black line corresponds to the perturbed individual. From left to right, we gradually increase the value of $\eta$ so as to step from the underdamped to the overdamped propagation regime (see \cref{par:dynamical_regimes}). In the left panel, the propagation is efficient and the flock is able to perform a collective turn. In the following two panels, the propagation is inefficient: the coherence of the flock is progressively lost, until the perturbed individual exits the flock without being followed. Videos corresponding to these different situations are included in the SM.
    The parameters used in the plots are $\chi = 1.25, J = 0.8, T = 8 \times 10^{-5}, A_0 = 30, N = 400, v_0 = 0.1$.}
    \label{fig:Edmondo}
\end{figure*}

\section{Open boundary conditions}
\label{par:openBC}
In the previous Section we investigated the model off-lattice but with periodic boundary conditions. In this case, as we discussed, the local perturbation is able to change the global direction of motion, even in the presence of strong dissipation. This is apparently at variance with experimental observations, where coherent turning is associated with underdamped inertial propagation. The reason for this mismatch is however related to the artificial nature of a box with periodic boundary conditions. Indeed, PBC can mask the role of dissipation: whenever the perturbed individual leaves the flock, it enters again from the opposite side of the box, where it can start spreading the signal again. Simulating the system with open boundary conditions (OBC), 
as we do in this Section,
is expected to eliminate this spurious effect.
This new setting mimics more closely the situation of real flocks and other groups, and will unravel the crucial role played by the underdamped propagation regime in allowing the flock to sustain a collective turn. 

\subsection{Role of the dynamical propagation regimes}
\label{sec:obc_regimes}
To implement the OBC, we 
first evolve the system off-lattice for a short time interval, until the scalar polarization reaches the stationary value $\Psi \sim 0.99$, and then we apply the field
to an individual placed in the middle of the group. We then proceed, as in the previous sections, to measure the evolution of the mean polarization angle. 
Since (now) the group has finite boundaries, more stringent conditions must be met in order for a global turn to take place. Indeed, the perturbed individual will quickly follow the field direction and deviate from the group's mean velocity: if all the other individuals do not follow (and turn by the {\it same} angle) soon enough, a finite difference in orientation with respect to the turn initiator will persist when the latter reaches the boundary. This can result either in the flock remaining compact but turning by a smaller amount, or else in the group's fragmentation. For a full turn to occur we therefore need  i) the information about the angular deviation to propagate intact to all individuals (\ie{} underdamped inertial propagation), and ii) the turn to be complete before the initiator hits the boundary (\ie{} not too large motility $v_0$). 

To verify this picture quantitatively, let us start by considering the first condition. We choose a small value of $v_0$ -- for which we know the analytical predictions to hold even off-lattice -- and we explore the various dynamical regimes of the model by appropriately varying the other parameters. Since in the off-lattice model large values of $\chi$ enhance clustering effects, we keep the inertia fixed and tune instead the dissipation $\eta$ to span the underdamped/overdamped spectrum. 


 The resulting curves for the mean polarization angle $\Phi(t)$ are reported in \cref{fig:OBC_phi_t}. This figure shows that only for sufficiently low $\eta$ (\ie{} in the underdamped inertial regime) the flock is able to sustain a collective turn and follow the perturbation. 
At higher values of $\eta$ (\ie{} in the overdamped dissipative regime), the dissipation is stronger and the perturbed individual leaves the flock before the whole group can turn, so that the global flight direction is only partially affected by the perturbation. By definition, the mean polarization angle $\Phi(t)$ saturates to $\alpha_p = \pi /2$ for $t \to \infty$ only if the entire flock remains coherent and aligns with the field, whereas $\Phi(\infty) < \pi /2$ corresponds to a fragmentation of the group. If the group splits into two components (one containing the perturbed individual and moving along the field direction, and another one going in another direction), then the resulting $\Phi(t)$ will be given by the weighted average over the two flight directions, with the weights represented by the sizes of the two clusters. The fewer individuals follow the perturbed one, the more $\Phi(\infty)$ will differ from $\pi/2$. 
In the inset of \cref{fig:OBC_phi_t} we thus plot the limiting polarization angle $\Phi(\infty)$ as a function of $\tau/\tau_s$, \ie{} the ratio between the dissipative and inertial timescales. As discussed in \cref{par:dynamical_regimes}, this ratio identifies whether the system is in the overdamped ($\tau/\tau_s<1$) or underdamped ($\tau/\tau_s>1$) regime of information propagation. This plot therefore shows that only when the system enters the underdamped inertial regime the turn becomes fully efficient and coherent (\ie{} $\Phi(\infty)=\pi/2$).

To further illustrate the mechanism described above, we also plot in \cref{fig:Edmondo} the single polarization angles  $\varphi_i(t)$ of all the individuals in the group, for a single realization of the turning event. At small values of $\eta$ in the underdamped regime, the flock remains coherent and the collective turn can take place. At intermediate values of $\eta$, only some individuals follow the perturbed one, and the group gets disrupted. Finally, at high $\eta$, the perturbed individual exits the flock while the flying direction of the rest of the group changes only slightly. Videos showing these turning events in real-time are included in the Supplementary Material (SM), supporting the scenario we just described.

\begin{figure*}
    \centering
    \includegraphics[width=\textwidth]{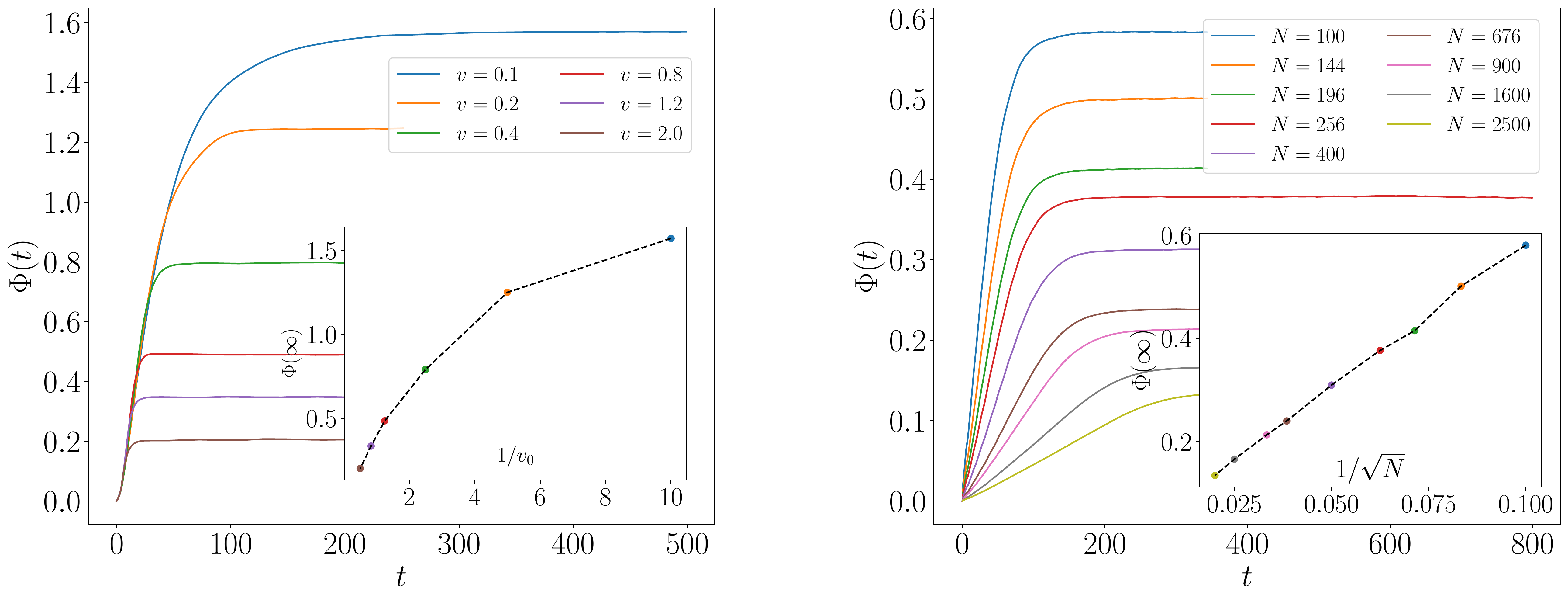}
    \put(-505,185){a)}
    \put(-235,185){b)}
    \caption{Off-lattice simulations,
    with open boundary conditions. Panel \textbf{(a)}: mean polarization angle $\Phi(t)$ for several activities $v_0 \in [0.1,2.0]$. Inset: long-time plateau $\Phi(t\to \infty)$ vs $1/v_0$ -- the dependence is approximately linear for large $v_0$, while $\Phi(t\to \infty)$ saturates to $\pi/2$ for smaller $v_0$ (see the main text). Panel \textbf{(b)}: mean polarization angle for several group sizes $N$. Inset: plateau $\Phi(t \to \infty)$ vs $1/ \sqrt{N}$, showing a linear dependence (as expected from \cref{eq:OBC_phi_t}). 
    The parameters used in both plots are $\chi = 1.25, J = 0.8, T = 8 \times 10^{-5}, A_0 = 30, n_\T{turns}=10$, while $ \eta = 0.3, N = 400$ in panel (a) and $\eta = 5, v_0 = 0.1$ in panel (b).}
    \label{fig:OBC}
\end{figure*}

\subsection{Role of the activity and dependence on group size}
Let us now discuss what is the role of the activity $v_0$ in determining the occurrence of a turn. To do so, we consider the system in its underdamped regime, where we know from the previous section that -- at small values of $v_0$ -- the system is able to fully reorient, following the perturbed individual, in the direction of the applied field. We thus choose in the simulation the same parameters as in \cref{fig:OBC_phi_t}, with $\eta = 0.3$. Then we start increasing the activity of the individuals, to see whether and how much the turning performance is affected.

As discussed before, the ability of the flock to sustain a collective turn is determined not only by the efficiency of information propagation, but also 
by the time taken by the perturbed individual to exit the group. We can expect that the higher $v_0$, the harder will be for the flock to turn, since the perturbed individual will leave the system sooner. A simple argument can be used to predict the final direction of flight $\Phi(\infty)$ of the flock depending on its activity. At the beginning of the turn the flock has large polarization $\Psi$, so we can imagine the system as if collectively moving straight; when an external step-like field is applied to a bird located in the middle of the flock, the perturbed individual changes its flight direction by $\pi /2$ within a negligible time span. 
The time taken by the perturbed bird to exit the flock is thus of the order of 
\begin{equation}
    t_\T{exit} \sim L / v_0 \, .
    \label{eq:exiting_time}
\end{equation}
In general, 
we can assume that the system stops turning at $t\sim t_\T{exit}$, so that $\Phi(\infty) \sim \Phi(t_\T{exit})$ (at least in the absence of oscillations). For the parameters considered in the simulation, the system is in the underdamped propagation regime but has a linear growth (see \cref{fig:OBC_phi_t}). We can adopt the on-lattice prediction in \cref{eq:Phi_pred_over} to estimate $\Phi(t_\T{exit})$, which gives (since $N=L^2$ and $\alpha_p=\pi/2$)
\begin{equation}
    \Phi(t=t_\T{exit}) \sim \dfrac{A_0}{N\eta} \dfrac{L}{v_0} = \dfrac{A_0}{\eta L v_0}\, .
    \label{eq:OBC_phi_t}
\end{equation}
Indeed, our current choice of parameters gives $v_0 \ll c_s\simeq 2.12 \ll v_0^{\text{lim}}$, so that the on-lattice estimate in \cref{eq:Phi_pred_over} is still reliable according to our discussion in \cref{eq:offlattice_role_dynamics}.
Equation~\eqref{eq:OBC_phi_t} suggests that, when $v_0$ increases, the final mean polarization angle should decrease. However, this estimate 
(and in particular the dependence on $1/v_0$) only hold when the exit time $t_\T{exit}$ falls within the linear growth regime of the polarization angle. This is not the case for sufficiently low activity $v_0$, because we expect the system to perform a full turn in this limit, and $\Phi(t=t_\T{exit})$ must therefore approach $\pi /2$.
This picture is confirmed by our numerical simulations. In \cref{fig:OBC}a we display the mean polarization angle for different values of the activity $v_0$, and we show that indeed for sufficiently large values of $v_0$ the group is unable to complete the turn. In the inset, we plot the endpoint $\Phi(t_\text{exit})$ as a function of $1/v_0$: as expected, the linear dependence predicted by \cref{eq:OBC_phi_t} gradually breaks down at small $v_0$, when $\Phi(t_\text{exit})$ approaches $\pi/2$.

It is interesting to compare \cref{eq:OBC_phi_t} with the response of the polarization $\Phi(t)$ to an impulse-like perturbation of finite duration $\cor{T}$. This is derived in \cref{app:wandering_goldstone} for the \textit{on-lattice} case with a calculation analogous to the one described in \cref{par:perturbation_result}, yielding
\begin{equation}
    \Phi(t\to \infty) = \frac{A_0 \sin\alpha_p  }{N \eta} \cor{T} \, .
    \label{eq:impulse_endpoint}
\end{equation}
The latter clearly agrees with the \textit{off-lattice} prediction in \cref{eq:OBC_phi_t} upon choosing $\cor{T}=t_\text{exit}$, \ie{} the time during which the perturbed individual remains inside the group (see \cref{eq:exiting_time}).

Expression (\ref{eq:OBC_phi_t}) also suggests that the final polarization angle, and therefore the efficiency with which the turn is performed, also depends on the size of the group itself. {\it Ceteris paribus}, large groups tend to be more in the overdamped regime, and they would take more time to perform a turn even on a fixed network. To better pinpoint these effects, we focus on the overdamped regime by using the same parameters as in \cref{fig:OBC_phi_t} with $\eta = 5$, and then we vary the size $N \in [100,1600]$. The result is displayed in \cref{fig:OBC}b, where the behaviour predicted by  \cref{eq:OBC_phi_t} is very well obeyed, including the linear dependence of the final flight direction on $1/ \sqrt{N}$ (see inset).

\section{Conclusions}
\label{par:conclusion}
Let us now summarize our results. The main aim of this work was to elucidate the interplay between size, motility and dynamical regimes, in the occurrence of perturbation-response events. More specifically, due to the strong functional relevance of anti-predatory response in finite groups, we focused on the ability of a group to change its global flight direction upon local perturbations. We modelled the perturbation in the form of an external field applied to a single individual of the group (see \cref{par:onlattice}). Even though the considered scheme is very simplified, it mimics quite reasonably many real instances of perturbations, from attacks of predators, to disturbances and obstacles -- which are typically perceived only by a subset of the group, but are eventually transmitted to other individuals thanks to mutual interactions. Experimental data indicate that, in the presence of local triggers, the way the group reacts can be different, ranging from the full coherent turns of starling flocks \cite{information_transfer} to the orientational cascades observed in fish schools \cite{Rosenthal_2015}. Our analysis helps to understand why this might occur. 

In the first part of this work (see \cref{par:onlattice}), we derived the dynamical response to local perturbations of the on-lattice system, and we tested the validity of our analytical expressions in numerical simulations. This allowed us to pinpoint the role of inertia and dissipation, both in the propagation mechanism of information and in the shape of a collective turn.
In \cref{par:offlattice}, we extended our analysis to the off-lattice system in the presence of periodic boundary conditions.  However, the PBC are extremely artificial when thinking of groups. More importantly, they hide the potentially disastrous effect of a strong dissipation by making a signal (the field), which is naturally finite in a motile finite group, long-standing. To overcome this problem, 
in \cref{par:openBC} we
considered systems with open boundary conditions. In this more realistic setup, we showed that for a coherent complete turn (collective change of direction) to occur, the group must i) live in the underdamped regime, so as to propagate information efficiently without damping, and ii) have a moderate motility, to avoid the `bullet' effect, when the individual which first perceives the perturbation leaves the group before all the others have the time to fully rearrange their direction -- see \cref{fig:Edmondo}.

In our analysis, we considered an external field which is turned on very rapidly, with a step-like profile, and perpendicularly to the original flight direction of the flock. In a way, this represents the worst possible scenario of an abrupt change, forcing the system (if it can) to exhibit a quick response. Of course, less dramatic situations might be possible, where the field changes slowly in time (\ie{} the parameter $\tau_\T{step}$ in \cref{eq:simil_step} is large), and/or of an angle smaller than $\pi/2$. In this case, we expect less stringent requirements for a collective response with OBC to occur. Indeed, the initiator changes its direction gradually due to the lesser strength of the field as compared to the social force of neighbours. This also implies that the initiator remains well inside the group for longer, giving more time to information to propagate and to other individuals to catch up.
It is possible that in this case the effect of dissipation 
may therefore be less dramatic \cite{private}. Future analysis of this aspect would certainly be interesting. However, given the variety and extension of possible perturbations occurring in realistic contexts, we believe that our analysis sets some benchmark criterion to understand efficient response behaviour.

\begin{acknowledgments}
We thank F. Ferretti, G. Pisegna and M. Viale for their support during the early stages of this work. We would like to thank A. Cavagna, T. Grigera and S. Ruffo for many interesting discussions. DV acknowledges support from MIUR PRIN project ``Coarse-grained description for non-equilibrium systems and transport phenomena (CO-NEST)'' n.~201798CZL. IG acknowledges support from MIUR PRIN project ``Response, control and learning: building new manipulation strategies in living and engineered active matter'' n.~2020PFCXPE.
\end{acknowledgments}

\appendix

\section{Derivation of the response}
\label{app:derivation_response}
In this Appendix we derive the response on-lattice of the polarization angle $\Phi(t)$ to a local perturbation, chosen in the form of an external field linearly coupled to one of the variables $\vb{v}_i$ (see \cref{par:eq_onlattice}). This will lead to the results anticipated in \cref{par:perturbation_result}. In the following, we mainly adopt the notation first introduced in \cite{Cavagna_2017}.

\subsection{Microscopic calculation}
\label{app:derivation_microscopic}
We start from the coupled Langevin equations~\eqref{eq:eqn_mot_SWA_1} and \eqref{eq:eqn_mot_SWA_2}, which we recast for convenience in the matrix form
\begin{equation}
    \dv{t} \va{\Psi} + \hat{L} \va{\Psi} = \va{F}(t) \, ,
    \label{eq:matrix_langevin}
\end{equation}
upon defining the the $2N$-dimensional vectors
\begin{equation}
    \va{\Psi} \equiv \mqty( \bm{\varphi} \\ \vb{s} ) \, , \;\;\; \va{F} \equiv \mqty( \vb{f}_{(\varphi)} \\ \vb{f}_{(s)})= \mqty( \vb{0} \\ \bm{\xi}+\vb{h}\sin\vb{\alpha} ) \, .
    \label{eq:bigvectors}
\end{equation}
In particular, the $N$-dimensional vector $\bm{\varphi}$ has components $\varphi_i(t)$.
We also introduced the block matrix
\begin{equation}
    \hat{L} = \mqty( \mathbb{O} & -\mathbb{I}/\chi \\ J \tilde{\Lambda} & \eta \, \mathbb{I}/\chi) \, ,
    \label{eq:L}
\end{equation}
where $\Tilde{\Lambda}$ is defined in terms of the discrete Laplacian $\Lambda_{ij}$ given in \cref{eq:discretelaplacian} as
\begin{equation}
\tilde{\Lambda} = \Lambda + \delta \Lambda \, , \;\;\;\;\;
(\delta \Lambda)_{ij} = \delta_{ij} \frac{h_i \cos\alpha_i}{J} \, .
\end{equation}
Equation~\eqref{eq:matrix_langevin} admits a formal solution as
\begin{equation}
    \va{\Psi} (t) = e^{-t\hat{L}}  \va{\Psi}(0) + \int_0^{t} \dd{t'} e^{-(t-t')\hat{L} }  \va{F}(t') \, ,
    \label{eq:formalsolution}
\end{equation}
which can be made explicit by diagonalizing $\hat{L}$.

\subsubsection{Spectrum of  $\hat{L}$}
Let us initially set $h_i=0$ in the expression for $\hat{L}$, which is equivalent to replacing $\tilde{\Lambda}$ with $\Lambda$ in \cref{eq:L}. We will first compute the unperturbed spectrum of $\hat{L}$ and only later add the effect of a small perturbing field $h_i$. Let us adopt for convenience the \textit{bra-ket} formalism familiar from Quantum Mechanics \cite{sakurai2017modern}, which allows us to express $\bm{\varphi} \rightarrow \ket{\varphi}$, $\vb{s} \rightarrow \ket{s}$, and
\begin{equation}
\va{\Psi} \rightarrow \ket{\Psi} = \ket{\varphi} \otimes \ket{s} \, ,
\end{equation}
where the symbol $\otimes$ denotes the tensor product operation. The components of $\ket{\varphi}$ and $\ket{s}$ along the \textit{position} basis
\begin{equation}
    \cor{B}_{pos}: \; \ket{i} \otimes \ket{i} \, ,
\end{equation}
where $\ket{i}$ is a $N$-dimensional vector, can be found by taking the scalar products
\begin{equation}
    \varphi^i = \braket{i}{\varphi} \, , \;\;\; s^i = \braket{i}{s}  \, .
\end{equation}
Conversely, let us denote as
\begin{equation}
    \cor{B}_{a}: \; \ket{a} \otimes \ket{a} \, , \;\;\;\;\; \Lambda \ket{a} = \lambda_a \ket{a}
\end{equation}
the basis in which $\Lambda$ is diagonal. The new components $\varphi^a$, labelled by the superscript $a$, are connected to the old ones ($\varphi^i$) via
\begin{align}
    \varphi^a &= \braket{a}{\varphi} = \bra{a} \sum_i \dyad{i} \ket{\varphi} \n\\
    &= \sum_i \braket{a}{i} \varphi^i \equiv \sum_i (U^{-1})\indices{^a_i}\, \varphi^i \, ,
\end{align}
where in the first line we inserted a decomposition of the identity $\mathbb{I}$, while in the second line we introduced the $N \times N$ matrix $U$, which enforces the change of basis
\begin{equation}
    U \, : \; \cor{B}_{a} \rightarrow \cor{B}_{pos} \, .
\end{equation}
When written in the basis $\cor{B}_{pos}$, the matrix $U$ contains the eigenvectors of $\Lambda$ as its columns; as a consequence, $U$ diagonalizes $\Lambda$ as
\begin{equation}
    \T{diag}(\lambda_1,\,\lambda_2,\,\dots) = U^{-1} \Lambda U \, .
\end{equation}
In the case of symmetric interactions $n_{ij}=n_{ji}$, the condition $\Lambda_{ij}=\Lambda_{ji}$ makes $U$ unitary ($U^{-1}=U^\dag $); note that the interaction network does not need to be regular (\eg{} a square lattice) in order for this to occur.

Working in the basis $\cor{B}_{a}$, we now look for the eigenvectors of $\hat{L}$:
\begin{equation}
    \hat{L}\ket{l} = \omega_l \ket{l}\, , \;\;\;\;\; \ket{l} = \ket{l}_\varphi \otimes \ket{l}_s \, .
    \label{eq:eigenvalueequation}
\end{equation}
Without loss of generality, we choose both $\ket{l}_\varphi$ and $\ket{l}_s$ to be proportional to $\ket{a}$, \ie{}
\begin{equation}
    \ket{l} = \ket{a} \otimes c(l)\ket{a} \;\;\; \xrightarrow{\text{$\cor{B}_a$}}\;\;\; \kappa \mqty(1 \\ c(l))\, ,
\end{equation}
where $\kappa$ is some normalization constant. The functional form of $c(l)$ has to be found by solving the eigenvalue equation \eqref{eq:eigenvalueequation}, which we can rephrase as
\begin{equation}
    \mqty(\omega_l & 1/\chi \\ - J \lambda_a & \omega_l - \eta/\chi) \mqty(1\\ c(l)) = \mqty(0\\ 0) \, .
    \label{eq:lineareigen}
\end{equation}
Solving for $c(l)$ leads to
\begin{equation}
-J \lambda_a - \omega_l^2 \chi + \eta \omega_l = 0 \, ,
\end{equation}
so that, calling as usual $\gamma=\eta/(2\chi)$, we obtain
\begin{equation}
    \omega_l = \gamma \pm i \omega_a \, , \;\;\;\;\;
    \omega_a \equiv \sqrt{J\lambda_a/\chi-\gamma^2} \, .
    \label{eq:omegadiscreto}
\end{equation}
Using \cref{eq:lineareigen} yields $c(l) = -\chi \omega_l$, which finally gives the eigenvectors
\begin{equation}
    \ket{l} \xrightarrow{\text{$\cor{B}_a$}} \kappa \mqty(1 \\ -\chi \omega_l ) \, .
\end{equation}
For each fixed value of $a$ (there are $N$ such values), \cref{eq:omegadiscreto} shows as expected a twofold degeneracy for $\omega_l \equiv \omega_\pm (\lambda_a)$. We can thus represent
\begin{equation}
    \hat{L} \xrightarrow[\text{$a$ fixed}]{\text{$\cor{B}_a$}} L \equiv \mqty(0 & -1/\chi \\  J \lambda_a & \eta/\chi) \, ,
\end{equation}
each of which is diagonalized by a matrix $M$ such that
\begin{equation}
    \T{diag}(\omega_+, \, \omega_-) = M^{-1}LM \, .
\end{equation}
The latter is non-unitary ($L$ is not symmetrical even when $\Lambda_{ij}=\Lambda_{ji}$), and reads
\begin{equation}
    M = \kappa \mqty( 1 & 1 \\ -\chi \omega_+ & -\chi \omega_-) , \; M^{-1} = \frac{1}{ 2 i \chi \kappa\omega_a} \mqty( -\chi \omega_- & -1 \\ \chi \omega_+ & 1) \, .
\end{equation}

\subsubsection{Temporal evolution}
Equation~\eqref{eq:formalsolution} takes a simple form in the basis $\cor{B}_a$. Indeed, by expressing
\begin{equation}
    \ket{\Psi} =  \sum_l \dyad{l} \ket{\Psi} = \sum_l \Psi^l \ket{l} \, , \;\;\; \ket{F} = \sum_l F^l \ket{l} 
\end{equation}
and by using the completeness relation
\begin{equation}
    e^{-t\hat{L}} = \sum_l e^{-t\omega_l} \dyad{l} \, ,
\end{equation}
one obtains the set of $2N$ decoupled equations
\begin{equation}
    \Psi^l (t) = e^{-t\omega_l}  \Psi^l(0) + \int_0^{t} \dd{t'} e^{-(t-t')\omega_l }  F^l(t') \, .
\label{eq:timeevolutionl}
\end{equation}
In order to translate this result back into the position basis, we need to:
\begin{enumerate}[(i)]
\item{Use the matrices $U$ and $U^{-1}$ on each of the two $N$-dimensional subspaces of the source term $\va{F}$ in \cref{eq:bigvectors}, so as to move to the basis in which the Laplacian is diagonal;}
\item{Apply the matrices $M$ and $M^{-1}$ on each $2$-dimensional subspace at fixed $a$, moving to the basis in which $L$ is diagonal. This gives the components $F^l(t)$, which contain the source terms;}
\item{Use \cref{eq:timeevolutionl} to compute the time evolution of each of the components $\Psi^l (t)$;}
\item{Apply the inverse of the two changes of basis used above (in reverse order), which finally gives $\varphi^i (t)$ and $s^i (t)$.}
\end{enumerate}
The above procedure is tedious but straightforward. Here we state the result for the variable $\varphi^i (t)$, 
\begin{align}
    \varphi^i (t) =& \sum_{a,j=1}^N U\indices{^i_a} (U^{-1})\indices{^a_j} \int_0^t \dd{t'} e^{-\gamma (t-t')} \times \n\\ 
    & \left\lbrace f^j_{(\varphi)}(t') \left[ \frac{\gamma}{\omega_a}\sin\omega_a(t-t') + \cos\omega_a(t-t') \right] \right. \n \\
    &+ \left. \frac{1}{\chi\omega_a} f^j_{(s)}(t') \sin\omega_a(t-t') \right\rbrace \, ,
    \label{eq:phi_i(t)}
\end{align}
where the second line vanishes in our case since $f^j_{(\varphi)}\equiv 0$. Equation~\eqref{eq:phi_i(t)} is still written in terms of the matrix $U$, whose columns are the normalized eigenvectors of the discrete Laplacian $\Lambda_{ij}$: as such, it is valid for any time-independent interaction network $n_{ij}$, whose eigenvectors could in principle even be obtained numerically. In the following, we will see that we do not need to compute explicitly these eigenvectors in order to derive an expression for $\Phi(t)$ in the case of symmetric interactions.

If we now specialize to the case considered in the main text, where only a field coupled to the phases $\varphi_i$ is considered, the term $f^j_{(\varphi)}$ is zero (see \cref{eq:ism_plan1-2,eq:ism_plan2-2}) and we get
\begin{align}
    \varphi^i (t) =& \sum_{a,j=1}^N U\indices{^i_a} (U^{-1})\indices{^a_j} \int_0^t \dd{t'} e^{-\gamma (t-t')} \times \n\\ 
    & \frac{1}{\chi\omega_a} [h_j(t')\sin\alpha_j+\xi_j(t')] \sin\omega_a(t-t') \, ,
    \label{eq:AKA_main_intermediate}
\end{align}
which corresponds to \cref{eq:main_intermediate} of the main text.

\subsubsection{Corrections to the spectrum of $\hat{L}$}
The first-order correction to the unperturbed spectrum when a small external field $\vb{h}$ is switched on can be computed by standard perturbation theory \cite{sakurai2017modern},
\begin{equation}
    \tilde{\lambda}_a = \lambda_a + \ev{\delta \Lambda}{a} \, , \;\;\;\;\;
    \ket{\tilde{a}} = \ket{a} + \sum_{b\neq a} \frac{\mel{b}{\delta \Lambda}{a}}{\lambda_a - \lambda_b} \ket{b} \, ,
\end{equation}
where $\lambda_a$ are the eigenvalues of the discrete Laplacian $\Lambda$ and $\ket{a}$ are its eigenvectors. This leads to
\begin{align}
    \tilde{\lambda}_a &= \lambda_a + \sum_{i} (U^{-1})\indices{^a_i} \frac{h_i\cos\alpha_i}{J} U\indices{^i_a} \, , \label{eq:correctedeigenvalues} \\
    \tilde{U}\indices{^i_a} &= U\indices{^i_a} + \sum_{b\neq a} \sum_{j} \frac{(U^\dag)\indices{^b_j} (h_j\cos\alpha_j) U\indices{^j_a}}{J(\lambda_a - \lambda_b)} U\indices{^i_b} \, ,
    \label{eq:correctedeigenvectors}
\end{align}
which can be used into \cref{eq:phi_i(t)} to replace $U\indices{^i_a}$ and $\lambda_a$, the latter being contained in $\omega_a$ via \cref{eq:omegadiscreto}.
Note that the corrections in \cref{eq:correctedeigenvalues,eq:correctedeigenvectors} vanish by choosing $\alpha_i\equiv \pi/2 \; \forall i$.

\subsubsection{Average polarization angle}
A remarkable property of the discrete Laplacian $\Lambda_{ij}$, which can be inferred from its definition in \cref{eq:discretelaplacian}, is
\begin{equation}
    \sum_j \Lambda_{ij} = \sum_j \left( -n_{ij} + \delta_{ij} \sum_k n_{ik} \right) =  0 \, ,
\end{equation}
which means that $\Lambda_{ij}$ always admits a zero mode (\ie{} a $\lambda_0=0$ eigenvalue) with constant right eigenvector. This implies that the matrix $U$, which contains the normalized eigenvectors as its columns, must be such that
\begin{equation}
    U\indices{^i_0} = 1/\sqrt{N} \;\;\; \forall i \, ,
\end{equation}
or equivalently $(U^{\dag})\indices{^0_i} = 1/\sqrt{N}$, since the transposed of the matrix $U$ contains the same (right) eigenvectors as its \textit{rows}. Note that in general these do \textit{not} coincide with the left eigenvectors of $\Lambda$: this is only the case when $\Lambda_{ij}$ is symmetric, because then the orthogonality condition reads
\begin{equation}
\mathbb{I} = U^{-1}U = U^\dag U \, .
\end{equation}
In the symmetric case we then have in particular
\begin{equation}
    \sum_i U\indices{^i_a} = \delta_{a0}\sqrt{N} \, ,
    \label{eq:perpendicularity}
\end{equation}
and summing over $i$ in \cref{eq:phi_i(t)} in order to obtain $\Phi(t)$, only the term with $a=0$ survives. Using \cref{eq:bigvectors} (or \cref{eq:AKA_main_intermediate}) with $\expval*{\bm{\xi}}=0$, we thus find
\begin{equation}
    \Phi (t) = \frac{1}{N}\sum_{j=1}^N \int_0^t \dd{t'} e^{-\gamma (t-t')} 
    \frac{ h_j(t')\sin\alpha_j}{\chi \omega_0} \sin \omega_0(t-t') \, ,
\label{eq:Phi_t_nonlocal}
\end{equation}
where we called $\omega_0 = \omega_{a=0}= i \gamma +\order{h}$ (see \cref{eq:omegadiscreto}).
Note that \cref{eq:Phi_t_nonlocal} is already of $\order{h}$, so that in general including the corrections to the spectrum of $\hat{L}$ due to the external field does not modify the leading-order contribution in perturbation theory.
The validity of this approximation extends to higher orders
by choosing $\alpha_j \equiv \pi/2$, because then we have seen that the first order corrections in \cref{eq:correctedeigenvalues,eq:correctedeigenvectors} vanish.
This is the choice we adopt in our numerical simulations.

By choosing a local field $h_i\propto \delta_{ip}=A_0\theta(t)$, we finally recover the expression for $\Phi(t)$ reported in the main text, \cref{eq:Phi_general}.
When $\alpha_p\neq \pi/2$, however, one has to be careful in some regions of the parameter space. Indeed, the full explicit expression of $\omega_0$ is (including corrections) $\omega_0=\sqrt{A_0 \cos\alpha/(N\chi)-\gamma^2} $. Usually the second term dominates, but for strong underdamping (\ie{} $\eta^2 N/(2\chi A_0\cos\alpha)\ll 1 $) it does not, and $\Phi(t)$ acquires a more complex structure.

\subsection{Coarse-grained calculation}
\label{app:derivation_coarsegrained}
The expression of the response $\varphi^i(t)$ we derived in \cref{eq:phi_i(t)} is general and it applies to any interaction network $n_{ij}$, even when it is not symmetric. However, we can get to \cref{eq:Phi_general} of the main text (which refers to the on-lattice case) with much less effort if we work in terms of the coarse-grained fields $\varphi(\vb{x},t)$ and $s(\vb{x},t)$.

The coarse-grained counterpart of our dynamical equations can be easily derived from \cref{eq:ism_plan1,eq:ism_plan2} and read \cite{tesi_venturelli}
\begin{align}
    \dot{\varphi} &= \fdv{\cor{U}}{s}  \, , \label{eq:canonical_phi}\\
    \dot{s} &= 
      - \fdv{\cor{U}}{{\varphi}}-
    \eta s + \xi \, , \label{eq:canonical_s}
\end{align}
where the continuous version of the potential is 
\begin{align}
    \cor{U} = \int \frac{\dd[d]{x}}{a^d}  \bigg[ & \frac{Jn_c a^2 }{2} \abs{\grad{\varphi}}^2  
    +\frac{1}{2}\cos\alpha\, h(\vb{x},t)\varphi^2    \n \\
    &- \sin\alpha\, h(\vb{x},t)\varphi \bigg]. 
    \label{eq:hamiltonian_coarsegrained}
\end{align}
Note that the component of the external field $\vb{h}$ parallel to the initial polarization $\vb{V}(t=0)$ plays the role of a \textit{mass} term for the field $\varphi$. We stress that $\cor{H}$ above has been derived under the SWA, hence it has to be considered as a low-temperature expansion of the ISM. In fact, a field theory can still be constructed if one relaxes this assumption, but it will in general be more complicated and possibly contain new, non-Gaussian terms \cite{3.99}.
%
From \cref{eq:canonical_s,eq:canonical_phi} we get
\begin{align}
    \dot{\varphi} &= 
    \frac{s}{\chi} \, , \\
    \dot{s} 
    &= Jn_ca^2 \laplacian{\varphi} - h \varphi \cos\alpha  + h \sin\alpha - \frac{\eta}{\chi}s  + \xi \, , \n
\end{align}
and deriving again the first equation we obtain
\begin{equation}
    \chi \ddot{\varphi} + \eta \dot{\varphi} -Jn_ca^2 \laplacian{\varphi} + h \varphi \cos\alpha  = h \sin\alpha + \xi \, .
\label{eq:ismcontinuum1}
\end{equation}
In analogy with the microscopic derivation, we now discard the term proportional to $h$ on the left-hand side, as we expect it to produce higher-order corrections for small $h$ (note that 
such term actually vanishes
for $\alpha=\pi/2$). This leads us to the linear problem
\begin{equation}
    \cor{D}_{\vb{x},t} \varphi(\vb{x},t) = f(\vb{x},t)
    \label{eq:familiarform}
\end{equation}
upon defining
\begin{align}
    \cor{D}_{\vb{x},t} &\equiv \chi \partial_{tt}^2 + \eta \partial_t - Jn_c a^2 \laplacian_{\vb{x}} \, , \label{eq:diff} \\ 
    f(\vb{x},t) &\equiv \sin\alpha\, h(\vb{x},t)  + \xi(\vb{x},t) \, .
\end{align}
The propagator of the differential operator in \cref{eq:diff} can be computed by standard methods \cite{Review_2018} to give, in the time-momentum domain,
\begin{equation}
    G(\vb{k},t) = e^{-\gamma t}\frac{\sin(\omega_{\vb{k}} t)}{\chi\omega_{\vb{k}}} \, ,
    \label{eq:propagator}
\end{equation}
where $\omega_{\vb{k}}\equiv c_s \sqrt{k^2-k_0^2}$ is precisely the one appearing on the right-hand side of \cref{eq:dispersion_relation}. This yields the average response of the system as
\begin{equation}
    \expval{\varphi(\vb{k},t)} = \int_0^t \dd{t'} \expval{ G(\vb{k},t-t')f(\vb{k},t') } \, .
    \label{eq:response_coarsegrained}
\end{equation}

We now turn to the polarization angle in \cref{eq:polarization_angle}, whose continuous-space version is
\begin{equation}
    \Phi(t) = \frac{1}{\cor{V}} \int_\cor{V} \dd[d]{x} \expval{\varphi(\vb{x},t)}  = \frac{1}{\cor{V}} \expval{\varphi(\vb{k}=\vb{0},t)} \, .
    \label{eq:def_total_pol_coarsegrained}
\end{equation}
Here we called $\cor{V}$ the volume occupied by the entire flock (\ie{} $\cor{V}\sim N$ on a square lattice), and we adopted the continuum prescription
\begin{equation}
    \sum_{\vb{k}} f(\vb{k}) \simeq \frac{\cor{V}}{(2\pi)^d} \int \dd[d]{k} f(\vb{k}) \, .
\end{equation}
Using \cref{eq:propagator,eq:response_coarsegrained} then simply gives
\begin{align}
    \Phi(t) &= \frac{1}{\cor{V}} \int_{-\infty}^t \dd{t'} G(\vb{k}=\vb{0},t-t') \sin\alpha H(\vb{k}=\vb{0},t') \n \\
    &= \frac{\sin\alpha}{\cor{V} \eta} \int_{-\infty}^t \dd{t'} H(\vb{k}=\vb{0},t') \left[ 1- e^{-2\gamma (t-t')}\right]\, .
\end{align}
To make contact with \cref{eq:Phi_general}, it is sufficient to choose a local field and give it a stepfunction-like time dependence,
\begin{equation}
    H(\vb{x},t) = A_0 \delta(\vb{x}-\vb{x}_0) \Theta(t)  \;\; \rightarrow \;\; H(\vb{k},t) = A_0 e^{i \vb{k} \cdot \vb{x}_0} \Theta(t) \, .
    \label{eq:local_perturbation}
\end{equation}
Let us note, however, the difference with respect to the microscopic approach of \cref{app:derivation_microscopic}: in that case, we have studied the effect of a perturbation applied on one of the microscopic variables $\vb{v}_i$. Here, conversely, the coarse-graining procedure has washed out the identity of the single microscopic degrees of freedom $\vb{v}_i$, and the perturbation in \cref{eq:local_perturbation} is thus applied locally at position $\vb{x}_0$. Moving off-lattice by introducing a large activity $v_0$ (see \cref{par:offlattice}), the former approach is expected to fail because of the time dependence in the interaction matrix $n_{ij}(t)$, while the latter approach fails both because the propagator $G(\vb{k},t)$ in \cref{eq:propagator} was computed on-lattice, and because it erroneously assumes that the perturbation remains fixed in space at $\vb{x}=\vb{x}_0$.

\section{Spin-wave decomposition and wandering of the order parameter}
\label{app:wandering}
In this Appendix we formally develop the spin-wave decomposition, and then we 
use it
to predict the \textit{wandering time} $\tau_w$ of the order parameter $\vb{V}$. As explained in \cref{par:perturbation_result}, the latter is defined as the persistence time of the total polarization $\vb{V}$ as it explores its broken-symmetry manifold under the effect of thermal fluctuations. The resulting estimate of $\tau_w$ is based on the on-lattice approximation, and it knows nothing about the activity $v_0$: one can thus interpret $\tau_w$ as the wandering time of the total magnetization in the XY model on a fixed lattice, and subject to an underdamped Langevin dynamics.

\subsection{Spin-wave decomposition}
\label{app:SWA}
When the system is in its ordered phase, it is useful to decompose each velocity vector into its components parallel and perpendicular to the global polarization $\vb{V}$,
\begin{equation}
    \vb{v}_i = v_i^L \vb{\hat{n}} + \bm{\pi}_i \, ,
    \label{eq:velocity_decomposition}
\end{equation}
where $\bm{\pi}_i$ is a $(d_v-1)$-dimensional vector, and by construction one has
\begin{equation}
    \sum_i \bm{\pi}_i = 0 \, .
    \label{eq:perp_condition}
\end{equation}
In the planar model, $\pi_i$ is simply a scalar,
\begin{equation}
    \pi_i = v_0 \sin\varphi_i \, .
\end{equation}
Now let us call $\vb{\hat{n}}$ the direction assumed by the polarization $\vb{V}$ at time $t=0$, but allow $\vb{V}(t)$ to change in time. We can decompose
\begin{equation}
\vb{V} = \frac{1}{N} \sum_i \vb{v}_i = (\vb{V}\cdot \vb{\hat{n}}) \vb{\hat{n}} + \vb{\delta V}^{\perp} \, ,
\end{equation}
where in the planar case
\begin{align}
    &(\vb{V}\cdot \vb{\hat{n}})(t) = \frac{1}{N} \sum_i v_i^L(t) \, , \label{eq:wanderingdef} \\
    &\vb{\delta V}^{\perp}(t) = \frac{1}{N} \sum_i \bm{\pi}_i = \frac{v_0}{N} \sum_i \sin\varphi_i(t)  \vb{\hat{\pi}} \simeq \frac{v_0 \vb{\hat{\pi}}}{N} \sum_i \varphi_i(t), 
\end{align}
the latter being valid within the SWA.

\subsection{Scalar polarization and mean angular deviation}
\label{app:fluctuations}
The SWA decomposition 
can be used to derive a relation between the scalar polarization, \ie{} $\Psi = \abs{\vb{V}}/v_0$, and the average fluctuation $\delta \varphi$ of a single velocity vector $\vb{v}_i$ around the global flight direction $\vb{V}$. Indeed, using \cref{eq:velocity_decomposition,eq:perp_condition} we have
\begin{equation}
    \Psi = \dfrac{1}{v_0 N} \Big| \sum_i \vb{v_i} \Big| = \dfrac{1}{ N} \Big| \sum_i \cos \varphi_i \Big| 
    \simeq 1 - \dfrac{1}{2N} \sum_i \varphi_i ^2 \, ,
    \label{eq:Psi_SWA}
\end{equation}
where in the last step we used the SWA.
Defining the mean fluctuation as 
\begin{equation}
    \delta \varphi \equiv \sqrt{\expval{\varphi^2}} = \dfrac{1}{N} \sum_i \varphi_i ^2 \, ,
\end{equation}
the relation stated in \cref{eq:pol_fluc} follows immediately from \cref{eq:Psi_SWA}.

\subsection{Persistence time $\tau_w$ of the order parameter}
We define the wandering time $\tau_w$ by the condition
\begin{equation}
    \expval{\abs{\vb{\delta \bm{\Psi}}^{\perp}(t\sim \tau_w)}^2} \sim \order{1} \, ,
    \label{eq:wandering_condition}
\end{equation}
where the brackets denote the average over the noise, and $\vb{V}=v_0\bm{\Psi}$. Using \cref{eq:wanderingdef}, we find under the SWA
\begin{align}
    \expval{\abs{\vb{\delta \bm{\Psi}}^{\perp}(t)}^2} &= \frac{1}{v_0^2 N^2} \sum_{ik} \expval{\bm{\pi}_i \cdot \bm{\pi}_k } \n\\
    &\simeq \frac{1}{N^2} \sum_{ik} \expval{\varphi^i(t)\varphi^k(t)} \, .
    \label{eq:wanderingavg}
\end{align}
The evolution of $\varphi^i(t)$ is given by \cref{eq:phi_i(t)} upon setting the external source field $\vb{h}=0$. This gives, for a \textit{symmetric} interaction network,
\begin{equation}
    \varphi^i (t) = \sum_{a,j=1}^N U\indices{^i_a} (U^{\dag})\indices{^a_j} \int_0^t \dd{t'} \frac{e^{-\gamma (t-t')}}{\chi\omega_a} \xi^j(t') \sin\omega_a(t-t'),
\end{equation}
where it is useful to isolate the contribution of the $a=0$ mode
\begin{equation}
\varphi^i (t) = \frac{1}{N} \sum_{j=1}^N \int_0^t \dd{t'} \frac{\xi^j(t')}{\eta} \left[ 1 - e^{-2\gamma (t-t')} \right] \; + \; (a\neq 0) \, .
\end{equation}
Indeed, thanks again to the property in \cref{eq:perpendicularity}, the modes $a\neq 0$ give no contribution; using the noise variance in \cref{eq:noise_corr} we thus obtain
\begin{equation}
    \expval{| \delta \bm{\Psi}^{\perp}(t)|^2} = \dfrac{2T}{N \eta } \left[t - \dfrac{ 1-e^{-2 \gamma t} }{\gamma} + \dfrac{1-e^{-4 \gamma t} }{4 \gamma} \right].
    \label{eq:wandering_origin}
\end{equation}
Specializing this expression for the two regimes $t \gg \gamma^{-1}$ and $t \ll \gamma^{-1}$ and using the condition in \cref{eq:wandering_condition}, we finally get
\begin{equation}
   \tau_w = 
   \begin{dcases}
   \frac{\eta N}{2 T} \, , & \T{for} \; t \gg \gamma^{-1} \, , \\
   \left(\frac{3}{16}\frac{N \eta}{T \gamma^2}\right)^{1/3} \, , & \T{for} \; t \ll \gamma^{-1} \, .
   \end{dcases}
   \label{eq:wandering_estimate}
\end{equation}
We stress again that \cref{eq:wandering_estimate} can be generically interpreted as the persistence time of the average polarization in the XY model subject to an underdamped Langevin dynamics. 
Its origin, as discussed above, is related to the presence of the zero modes of the Laplacian, \ie{} to the original rotational invariance of the potential function $\cor{U}$ of the velocities (see \cref{eq:heisenberg}). This is a common feature of $O(n)$ models, and it is independent of the kind of dynamics adopted to let the system evolve. Indeed, the wandering of the order parameter and the persistence time have been computed in a variety of works, see \eg{} \cite{Leoncini_98,Lepri_2001} where a microcanonical dynamics for the XY model was considered, or the more recent \cite{Cavagna_2017,Ginelli_2022} where the ISM and the Vicsek model were analyzed.

\section{Symmetry breaking in 2d spin systems}
\label{app:mermin_wagner}
In this Appendix we address the issue of the presence of a spontaneous magnetization in the XY model (to which our system effectively reduces when the activity $v_0$ is set to zero). We start by recalling a heuristic argument which is standard in statistical mechanics \cite{LeBellac}; a more rigorous proof due to Mermin and Wagner can be found in \cite{merminwagner}. Let us call $d$ the dimension of the physical space and let $n$ be the dimension of the order parameter (in our case $d=n=2$). 

Suppose that a nonzero spontaneous magnetization exists near $T=0$, and let us inspect the stability of the corresponding ordered state against small thermal fluctuations.
At low temperature, we can assume all the spins to be almost aligned in one direction, so that we can work within the continuum limit and the SWA. We thus consider a Hamiltonian of the form
\begin{equation}
    \mathcal{H} = \dfrac{\cor{J}}{2}  \int d^2 x \sum_{\alpha=1}^{n-1} [\nabla \varphi_{\alpha}(\vb{x})]^2 \, ,
    \label{eq:hamilt_XY}
\end{equation} 
corresponding to a Gaussian and massless field theory whose Fourier-space propagator reads \cite{LeBellac}
\begin{equation}
    \label{eq:gauss_prop}
    \widetilde{G}_0(\vb{k}) = T/(\cor{J} \vb{k}^2) \, .
\end{equation}
The fluctuation $\Delta$ of the order parameter can be estimated as
\begin{align}
    \label{eq:fluctuations_xy}
    \Delta &\equiv \sum_{\alpha=1}^{n-1} \expval{\varphi_{\alpha} ^2 (\vb{x})}= (n-1) \expval{\varphi_1^2 (\vb{x})} \\  &= (n-1) G_0(\vb{r}=0) 
    =\dfrac{(n-1)T}{\cor{J}} \int_{\pi/L}^{\pi/a}  \dfrac{\dd[d]{k}}{(2\pi)^d \, \vb{k}^2} \, , \n
\end{align} 
where the integration cutoffs are given by the system size $L$ and the lattice spacing $a$. For $d>2$ the integral in \cref{eq:fluctuations_xy} is infrared-convergent, so that $\Delta \to 0$ as $T\to 0$, which is consistent with our initial assumption that the ordered state is stable against fluctuations. However, in $d=2$ one has
\begin{equation}
    \Delta \sim \dfrac{(n-1)T}{\cor{J}} \ln{\dfrac{L}{a}},
\end{equation} 
showing that $\Delta \to \infty$ when $L \to \infty$: thus, the long-wavelength fluctuations destabilize the long-range order in the thermodynamic limit. In fact, it is well-known that in $d=2$ a phase transition is observed at $T=T_\T{BKT}$ \cite{Kosterlitz_1973}: for $T<T_\T{BKT}$ the correlation functions exhibit a scale-free decay, but the average magnetization remains zero (this is known as \textit{quasi long-range order}).

This classical argument explains the absence of long-range order in $d \leq 2$ continuous systems in the thermodynamic limit. Conversely, finite-size systems do exhibit a continuous transition at a critical temperature $T_c$, so that below $T_c$ we observe a nonzero spontaneous magnetization \cite{Bramwell_1993,Roomany_80,Chung_99,Leoncini_98,Lepri_2001}. Of course the value of the magnetization (slowly) decays as an inverse power of the system size $N$, so that there is no contradiction with the BKT theory: a low-temperature, spin-wave analysis renders a total magnetization \cite{Chester_1979,Bramwell_1993}
\begin{equation}
    M(N,T) = \left( \frac{1}{2N} \right)^{T/(8\pi \cor{J})}  ,
    \label{eq:M_N}
\end{equation}
hence for instance $M\sim N^{-1/16}$ at the BKT transition. In fact, one finds $M\sim \order{1}$ for a system with $N\sim \order{10^2-10^3}$ like the ones we analyze in this work: this is why in \cref{par:onlattice} we can still study the global order parameter $\vb{V}$ even within the on-lattice approximation. On the other hand, in the off-lattice case the coarse-graining procedure which led to \cref{eq:hamilt_XY} breaks down, so that the argument above does not apply -- indeed, spontaneous symmetry breaking and long-range order are well-known to take place in $d=2$ active systems \cite{Toner_1998}.

Finally, we can however ask how the computations performed in this work relate to the stability of order and the Mermin-Wagner result \cite{merminwagner}. In our numerical analysis, as stated above, we considered sizes and temperatures for which the magnetization/polarization remains of $\order{1}$ according to \cref{eq:M_N}. What would happen, though, if we considered larger sizes? If we prepare the system in an ordered state and then apply a field, the polarization angle will follow exactly the same behaviour as described in the main text, provided that the time scale of the turn is sufficiently fast. Note that actually \cref{eq:response} does not depend on the physical dimension $d$ of the space where the system lives. However, there are other quantities that crucially depend on the space dimension: the scalar polarization $\Psi$ is one of them. Once the turn is complete and the instantaneous mean polarization is along the direction of the field, we can compute the scalar polarization within the SWA from \cref{eq:Psi_SWA}, assuming as reference direction the one of the field and using the expression for $\varphi_i(t)$ derived in \cref{app:derivation_response}. It is easy to see that the computation gives back asymptotically the equilibrium estimate, meaning that -- once the field is set back to zero -- the polarization in $d=2$ is destroyed by fluctuations, while in larger dimensions it remains finite.

\section{Numerical integration scheme}
\label{app:numerical_integration}
We start by introducing $\sigma \equiv \sqrt{2T\eta}$ and
\begin{equation}
    f_i(\vb{x}(t),t) \equiv J \sum_{j} n_{ij}(t) \sin(x_j(t) - x_i(t)) \, ,
    \label{eq:force_numerical}
\end{equation}
so that the laws of motion 
in \cref{eq:ism_planr,eq:ism_plan1-2,eq:ism_plan2-2}
can be expressed as
\begin{align}
    \dd{\varphi_i(t)} =&\, \chi^{-1} s_i(t) \dd{t} \, , \\
    \dd{s_i(t)} =&\, \left[ f_i(\bm{\varphi}(t),t) -2\gamma s_i(t) - h_i(t) \sin \varphi_i(t) \right] \dd{t} \n\\
    &+ \sigma \dd{W_i(t)} \, ,
    \label{eq:systemnumerical}
\end{align}
where $W_i(t)$ is a Wiener process satisfying $\expval{\vb{W}(t)} = \vb{0}$ and
\begin{equation}
    \expval{W_i(t) W_j(t')} = \delta_{ij} \min \left( t,t' \right) \, .
\label{app:wienerdef}
\end{equation}
We aim at discretizing these stochastic differential equations in order to obtain a numerical integration scheme which is at least of the second order in the integration timestep $\Delta t$. To this end, we generalize the procedure described in \cite{ciccotti}
to the case in which $f_i(\vb{x}(t),t)$ can depend explicitly on time $t$. A lengthy but straightforward calculation \cite{tesi_venturelli} leads to
\begin{align}
    \varphi_i(t+\Delta t) =&\, \varphi_i(t) + s_i(t) \Delta t /\chi + A_i(t) + \order{\Delta t^{5/2}} \, , \\
    s_i(t+\Delta t) =&\, s_i(t) + \Delta t \left[ \bar{f}_i - 2\gamma s_i(t) - \overline{h_i \sin \varphi}_i \right] \n\\
    &+ \sigma \sqrt{\Delta t} \xi_i  -2\gamma A_i(t) + \order{\Delta t^{5/2}} \, , \label{eq:sistemanumerico}
\end{align}
where the bar over a variable stands for its midpoint value:
\begin{equation}
    \bar{f}_i = \frac{1}{2} \left[ f_i(\bm{\varphi}(t+\Delta t),t+\Delta t) + f_i(\bm{\varphi}(t),t) \right] \, .
\end{equation}
We also introduced the auxiliary variable
\begin{align}
    A_i(t) \equiv&  \left[ f_i(\bm{\varphi}(t),t) - 2\gamma  s_i(t) - h_i(t) \sin \varphi_i(t) \right] \Delta t^2/2 \n\\
    &+ \left( \xi_i + \zeta_i/\sqrt{3} \right) \sigma \Delta t^{3/2}/2 \, ,
\end{align}
where $\xi_i$ and $\zeta_i$ are white uncorrelated Gaussian variables with zero mean and unit variance.

There remains to specify how the positions $\vb{r}_i(t)$ evolve according to their corresponding velocity $\vb{v}_i(t)$. A simple choice is the Euler-Cromer update rule \cite{frenkel,tuckerman},
\begin{equation}
    \vb{r}(t+\Delta t)  = \vb{r}(t) + \vb{v}(t+\Delta t)\Delta t \, ,
\end{equation}
where each component of $\vb{r}_i$ has to be recast in $[0,L]$ in the case of periodic boundary conditions.
The connectivity matrix $n_{ij}(t)$ contained in the force term $\vb{f}(\vb{x}(t),t)$ in \cref{eq:force_numerical} must also be updated regularly, so as to take into account the relative motion of the individual positions. This is obtained by the \textit{cell-list} method \cite{frenkel,tuckerman}, which consists of dividing the lattice into cells, and assigning to each particle a label indicating the cell it occupies at a given time $t$; at the same time, we associate to each cell the list of the current occupants. This expedient speeds up the computation of the interacting force contributions, which become of $\order{N}$ (rather than $\order{N^2}$) since we only have to cycle over particles belonging to the same or neighbouring cells, and only then check if their distance lies below the metric interaction radius (see \cref{par:vicsek}).

The complete code used for numerical simulations (written in C) is available open source here \cite{code}.


\begin{figure}[t]
\centering
\includegraphics[width=\columnwidth]{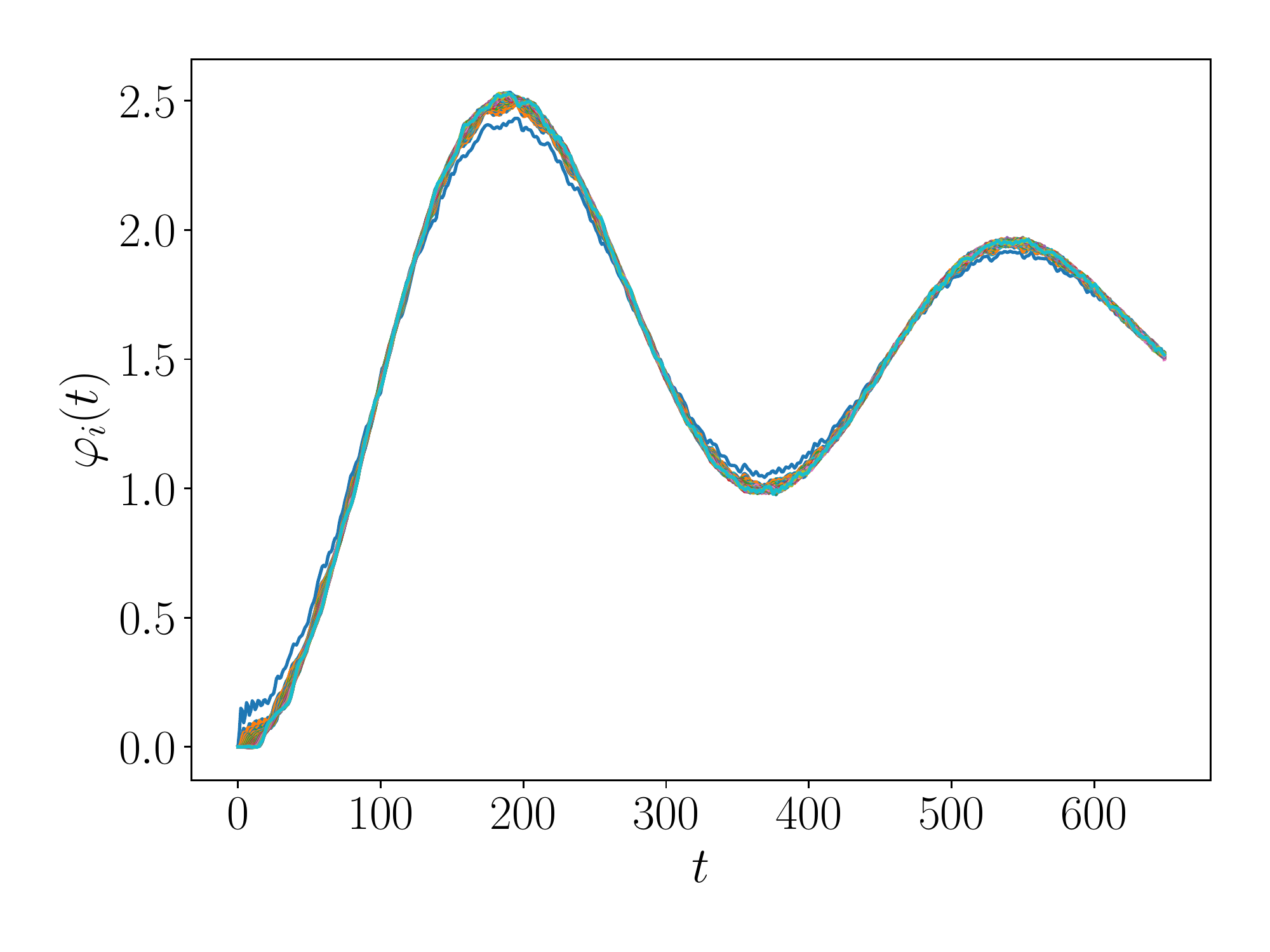}
\caption{Short-time evolution of $\varphi_i(t)$ for various individuals, in the deeply underdamped regime (on-lattice case). The individual phases $\varphi_i$ evolve coherently together, hence justifying the use of the mean-field approximation in studying global damped oscillations, as we did in \cref{app:oscillations}. We used $\chi=200, A_0=30, J=50, T=0.005, \eta =1, N=400$.}
\label{fig:meanfield_all}
\end{figure}

\section{Oscillations of the polarization angle in the underdamped regime}
\label{app:oscillations}
We have noted in \cref{par:collective} how, in the deeply underdamped regime, the mean polarization angle $\Phi(t)$ may occasionally overshoot the final angle $\alpha$ imposed by the external perturbation, and perform underdamped oscillations around the latter. An intuitive understanding of this phenomenon can be grasped by adopting again the coarse-grained description of \cref{app:derivation_coarsegrained}. We start from \cref{eq:ismcontinuum1} and set $\alpha=0$, which corresponds to a spin-wave approximation around the \textit{final} direction reached by the flock at the end of the turn. We additionally apply the mean-field approximation $\varphi(\vb{x},t)=\varphi_0(t)$, whose validity can be checked \textit{a posteriori}, thus obtaining
\begin{equation}
        \chi \ddot{\varphi}_0 + \eta \dot{\varphi}_0 + h \varphi_0 = \xi(t) \, .
\end{equation}
We now choose the external field as $h(\vb{x},t) = A_0 \delta(\vb{x}) \Theta(t)$.
Using the definition of $\Phi(t)$ in \cref{eq:def_total_pol_coarsegrained} and taking the average over the noise, we obtain an evolution equation for the total polarization in the form
\begin{equation}
        \ddot{\Phi} + \frac{\eta}{\chi} \dot{\Phi} + \frac{A_0}{\chi N} \Phi = 0 \, .
\end{equation}
This way we recover the equation of motion of a simple harmonic oscillator with damping coefficient $\gamma=\eta/(2\chi)$ and frequency $\Omega=\sqrt{A_0/(\chi N)}$.
%

Note that using the mean-field framework means assuming all the individuals to turn coherently, as if the system behaved as a rigid body. Hence, we are neglecting the site-to-site signal propagation which occurs within the flock at finite speed $c_s$ -- see \cref{par:propagationlaw}. However, the latter can generally be ignored when studying the global underdamped oscillations, since their associated timescale $\Omega^{-1}$ is much larger than the typical time taken by the information to travel from one individual to another, \ie{} $\Omega^{-1} \gg a/ c_s$, where $a$ is the lattice spacing. We check this explicitly in \cref{fig:meanfield_all}, where we plot $\varphi_i(t)$ for various individuals, and observe that the corresponding curves almost coincide.

\section{Impulse-like perturbations}
\label{app:wandering_goldstone}
In this Appendix 
we derive the response of the system to a finite-duration perturbation, which provides useful insights for the OBC case analyzed in \cref{par:openBC}. 
Let us then specialize the response in \cref{eq:Phi_general} to the case in which the local external field $\vb{h}_p(t)$ has an impulse-like time dependence of the form
\begin{equation}
    h_p(t) = A_0 \Theta(t) \Theta(\cor{T} -t) \, ,
\end{equation}
where $\cor{T}$ is the impulse duration. This leads to 
\begin{align}
    \Phi(t) = \frac{A_0 \sin\alpha_p  }{N \eta} \left[  \cor{T} + \frac{e^{-2\gamma t}}{2\gamma} \left( 1- e^{2\gamma \cor{T}} \right) \right] \, ,
    \label{eq:response_impulse}
\end{align}
valid as long as $\Phi(t) \ll \alpha_p$. In the limit $N \to \infty$, the long-$t$ value of the response in \cref{eq:response_impulse} vanishes, 
showing (as expected) that a \textit{local} and \textit{impulse-like} perturbation has no influence on the system in the thermodynamic limit. 

Curiously, \cref{eq:response_impulse} also implies that the endpoint $\Phi(t\to \infty)$ is the one given in \cref{eq:impulse_endpoint},
which does \textit{not} depend on the rotational inertia $\chi$. This may appear counter-intuitive: how can a finite external perturbation win a possibly very large rotational inertia? To answer this point, it is sufficient to note that a system with large inertia $\chi$ will keep rotating even after we stop applying the external \textit{torque}. Indeed,
consider the analogous problem of a point particle $x(t)$ of \textit{mass} $\chi$ subject to viscous friction,
\begin{equation}
    \chi \ddot x = - \eta \dot x \, . 
\end{equation}
By the impulse-momentum theorem, applying an impulse $F \cdot \cor{T}$ on the particle initially at rest changes its velocity from $\dot x=0$ to $\dot x =F \cor{T}/\chi $. It is then a simple exercise to check that the endpoint $x(t \to \infty)$ of the particle trajectory is actually independent of its mass $\chi$.







\bibliography{references}

\begin{thebibliography}{44}%
\makeatletter
\providecommand \@ifxundefined [1]{%
 \@ifx{#1\undefined}
}%
\providecommand \@ifnum [1]{%
 \ifnum #1\expandafter \@firstoftwo
 \else \expandafter \@secondoftwo
 \fi
}%
\providecommand \@ifx [1]{%
 \ifx #1\expandafter \@firstoftwo
 \else \expandafter \@secondoftwo
 \fi
}%
\providecommand \natexlab [1]{#1}%
\providecommand \enquote  [1]{``#1''}%
\providecommand \bibnamefont  [1]{#1}%
\providecommand \bibfnamefont [1]{#1}%
\providecommand \citenamefont [1]{#1}%
\providecommand \href@noop [0]{\@secondoftwo}%
\providecommand \href [0]{\begingroup \@sanitize@url \@href}%
\providecommand \@href[1]{\@@startlink{#1}\@@href}%
\providecommand \@@href[1]{\endgroup#1\@@endlink}%
\providecommand \@sanitize@url [0]{\catcode `\\12\catcode `\$12\catcode
  `\&12\catcode `\#12\catcode `\^12\catcode `\_12\catcode `\%12\relax}%
\providecommand \@@startlink[1]{}%
\providecommand \@@endlink[0]{}%
\providecommand \url  [0]{\begingroup\@sanitize@url \@url }%
\providecommand \@url [1]{\endgroup\@href {#1}{\urlprefix }}%
\providecommand \urlprefix  [0]{URL }%
\providecommand \Eprint [0]{\href }%
\providecommand \doibase [0]{https://doi.org/}%
\providecommand \selectlanguage [0]{\@gobble}%
\providecommand \bibinfo  [0]{\@secondoftwo}%
\providecommand \bibfield  [0]{\@secondoftwo}%
\providecommand \translation [1]{[#1]}%
\providecommand \BibitemOpen [0]{}%
\providecommand \bibitemStop [0]{}%
\providecommand \bibitemNoStop [0]{.\EOS\space}%
\providecommand \EOS [0]{\spacefactor3000\relax}%
\providecommand \BibitemShut  [1]{\csname bibitem#1\endcsname}%
\let\auto@bib@innerbib\@empty
\bibitem [{\citenamefont {Lima}(1995)}]{LIMA_1995}%
  \BibitemOpen
  \bibfield  {author} {\bibinfo {author} {\bibfnamefont {S.~L.}\ \bibnamefont
  {Lima}},\ }\bibfield  {title} {\bibinfo {title} {Back to the basics of
  anti-predatory vigilance: the group-size effect},\ }\href
  {https://doi.org/https://doi.org/10.1016/0003-3472(95)80149-9} {\bibfield
  {journal} {\bibinfo  {journal} {Anim. Behav.}\ }\textbf {\bibinfo {volume}
  {49}},\ \bibinfo {pages} {11} (\bibinfo {year} {1995})}\BibitemShut {NoStop}%
\bibitem [{\citenamefont {Krause}\ \emph {et~al.}(2002)\citenamefont {Krause},
  \citenamefont {Krause}, \citenamefont {Ruxton},\ and\ \citenamefont
  {Ruxton}}]{Krause2002living}%
  \BibitemOpen
  \bibfield  {author} {\bibinfo {author} {\bibfnamefont {J.}~\bibnamefont
  {Krause}}, \bibinfo {author} {\bibfnamefont {P.}~\bibnamefont {Krause}},
  \bibinfo {author} {\bibfnamefont {G.}~\bibnamefont {Ruxton}},\ and\ \bibinfo
  {author} {\bibfnamefont {G.}~\bibnamefont {Ruxton}},\ }\href
  {https://books.google.it/books?id=HAoUFfVFtMcC} {\emph {\bibinfo {title}
  {Living in Groups}}},\ Oxford Series in Ecology and Evolution\ (\bibinfo
  {publisher} {OUP Oxford},\ \bibinfo {year} {2002})\BibitemShut {NoStop}%
\bibitem [{\citenamefont {Katz}\ \emph {et~al.}(2011)\citenamefont {Katz},
  \citenamefont {Tunstrøm}, \citenamefont {Ioannou}, \citenamefont {Huepe},\
  and\ \citenamefont {Couzin}}]{Couzin_2011}%
  \BibitemOpen
  \bibfield  {author} {\bibinfo {author} {\bibfnamefont {Y.}~\bibnamefont
  {Katz}}, \bibinfo {author} {\bibfnamefont {K.}~\bibnamefont {Tunstrøm}},
  \bibinfo {author} {\bibfnamefont {C.~C.}\ \bibnamefont {Ioannou}}, \bibinfo
  {author} {\bibfnamefont {C.}~\bibnamefont {Huepe}},\ and\ \bibinfo {author}
  {\bibfnamefont {I.~D.}\ \bibnamefont {Couzin}},\ }\bibfield  {title}
  {\bibinfo {title} {Inferring the structure and dynamics of interactions in
  schooling fish},\ }\href {https://doi.org/10.1073/pnas.1107583108} {\bibfield
   {journal} {\bibinfo  {journal} {Proc. Natl. Acad. Sci.}\ }\textbf {\bibinfo
  {volume} {108}},\ \bibinfo {pages} {18720} (\bibinfo {year}
  {2011})}\BibitemShut {NoStop}%
\bibitem [{\citenamefont {Rosenthal}\ \emph {et~al.}(2015)\citenamefont
  {Rosenthal}, \citenamefont {Twomey}, \citenamefont {Hartnett}, \citenamefont
  {Wu},\ and\ \citenamefont {Couzin}}]{Rosenthal_2015}%
  \BibitemOpen
  \bibfield  {author} {\bibinfo {author} {\bibfnamefont {S.~B.}\ \bibnamefont
  {Rosenthal}}, \bibinfo {author} {\bibfnamefont {C.~R.}\ \bibnamefont
  {Twomey}}, \bibinfo {author} {\bibfnamefont {A.~T.}\ \bibnamefont
  {Hartnett}}, \bibinfo {author} {\bibfnamefont {H.~S.}\ \bibnamefont {Wu}},\
  and\ \bibinfo {author} {\bibfnamefont {I.~D.}\ \bibnamefont {Couzin}},\
  }\bibfield  {title} {\bibinfo {title} {Revealing the hidden networks of
  interaction in mobile animal groups allows prediction of complex behavioral
  contagion},\ }\href {https://doi.org/10.1073/pnas.1420068112} {\bibfield
  {journal} {\bibinfo  {journal} {Proc. Natl. Acad. Sci.}\ }\textbf {\bibinfo
  {volume} {112}},\ \bibinfo {pages} {4690} (\bibinfo {year}
  {2015})}\BibitemShut {NoStop}%
\bibitem [{\citenamefont {Handegard}\ \emph {et~al.}(2012)\citenamefont
  {Handegard}, \citenamefont {Boswell}, \citenamefont {Ioannou}, \citenamefont
  {Leblanc}, \citenamefont {Tj{\o}stheim},\ and\ \citenamefont
  {Couzin}}]{Handegard2012}%
  \BibitemOpen
  \bibfield  {author} {\bibinfo {author} {\bibfnamefont {N.~O.}\ \bibnamefont
  {Handegard}}, \bibinfo {author} {\bibfnamefont {K.~M.}\ \bibnamefont
  {Boswell}}, \bibinfo {author} {\bibfnamefont {C.~C.}\ \bibnamefont
  {Ioannou}}, \bibinfo {author} {\bibfnamefont {S.~P.}\ \bibnamefont
  {Leblanc}}, \bibinfo {author} {\bibfnamefont {D.~B.}\ \bibnamefont
  {Tj{\o}stheim}},\ and\ \bibinfo {author} {\bibfnamefont {I.~D.}\ \bibnamefont
  {Couzin}},\ }\bibfield  {title} {\bibinfo {title} {The dynamics of
  coordinated group hunting and collective information transfer among schooling
  prey},\ }\href {https://doi.org/10.1016/j.cub.2012.04.050} {\bibfield
  {journal} {\bibinfo  {journal} {Curr. Biol.}\ }\textbf {\bibinfo {volume}
  {22}},\ \bibinfo {pages} {1213} (\bibinfo {year} {2012})}\BibitemShut
  {NoStop}%
\bibitem [{\citenamefont {Herbert-Read}\ \emph {et~al.}(2015)\citenamefont
  {Herbert-Read}, \citenamefont {Buhl}, \citenamefont {Hu}, \citenamefont
  {Ward},\ and\ \citenamefont {Sumpter}}]{HerbertRead2015}%
  \BibitemOpen
  \bibfield  {author} {\bibinfo {author} {\bibfnamefont {J.~E.}\ \bibnamefont
  {Herbert-Read}}, \bibinfo {author} {\bibfnamefont {J.}~\bibnamefont {Buhl}},
  \bibinfo {author} {\bibfnamefont {F.}~\bibnamefont {Hu}}, \bibinfo {author}
  {\bibfnamefont {A.~J.~W.}\ \bibnamefont {Ward}},\ and\ \bibinfo {author}
  {\bibfnamefont {D.~J.~T.}\ \bibnamefont {Sumpter}},\ }\bibfield  {title}
  {\bibinfo {title} {Initiation and spread of escape waves within animal
  groups},\ }\href {https://doi.org/10.1098/rsos.140355} {\bibfield  {journal}
  {\bibinfo  {journal} {Roy. Soc. Open Sci.}\ }\textbf {\bibinfo {volume}
  {2}},\ \bibinfo {pages} {140355} (\bibinfo {year} {2015})}\BibitemShut
  {NoStop}%
\bibitem [{\citenamefont {Calovi}\ \emph {et~al.}(2015)\citenamefont {Calovi},
  \citenamefont {Lopez}, \citenamefont {Schuhmacher}, \citenamefont {Chaté},
  \citenamefont {Sire},\ and\ \citenamefont {Theraulaz}}]{Calovi_2015}%
  \BibitemOpen
  \bibfield  {author} {\bibinfo {author} {\bibfnamefont {D.~S.}\ \bibnamefont
  {Calovi}}, \bibinfo {author} {\bibfnamefont {U.}~\bibnamefont {Lopez}},
  \bibinfo {author} {\bibfnamefont {P.}~\bibnamefont {Schuhmacher}}, \bibinfo
  {author} {\bibfnamefont {H.}~\bibnamefont {Chaté}}, \bibinfo {author}
  {\bibfnamefont {C.}~\bibnamefont {Sire}},\ and\ \bibinfo {author}
  {\bibfnamefont {G.}~\bibnamefont {Theraulaz}},\ }\bibfield  {title} {\bibinfo
  {title} {Collective response to perturbations in a data-driven fish school
  model},\ }\href {https://doi.org/10.1098/rsif.2014.1362} {\bibfield
  {journal} {\bibinfo  {journal} {J. R. Soc. Interface}\ }\textbf {\bibinfo
  {volume} {12}},\ \bibinfo {pages} {20141362} (\bibinfo {year}
  {2015})}\BibitemShut {NoStop}%
\bibitem [{\citenamefont {Klamser}\ and\ \citenamefont
  {Romanczuk}(2021)}]{Klamser2021}%
  \BibitemOpen
  \bibfield  {author} {\bibinfo {author} {\bibfnamefont {P.~P.}\ \bibnamefont
  {Klamser}}\ and\ \bibinfo {author} {\bibfnamefont {P.}~\bibnamefont
  {Romanczuk}},\ }\bibfield  {title} {\bibinfo {title} {Collective predator
  evasion: Putting the criticality hypothesis to the test},\ }\href
  {https://doi.org/10.1371/journal.pcbi.1008832} {\bibfield  {journal}
  {\bibinfo  {journal} {{PLOS} Comput. Biol.}\ }\textbf {\bibinfo {volume}
  {17}},\ \bibinfo {pages} {e1008832} (\bibinfo {year} {2021})}\BibitemShut
  {NoStop}%
\bibitem [{\citenamefont {Attanasi}\ \emph {et~al.}(2014)\citenamefont
  {Attanasi}, \citenamefont {Cavagna}, \citenamefont {Castello}, \citenamefont
  {Giardina}, \citenamefont {Grigera}, \citenamefont {Jeli{\'{c}}},
  \citenamefont {Melillo}, \citenamefont {Parisi}, \citenamefont {Pohl},
  \citenamefont {Shen},\ and\ \citenamefont {Viale}}]{information_transfer}%
  \BibitemOpen
  \bibfield  {author} {\bibinfo {author} {\bibfnamefont {A.}~\bibnamefont
  {Attanasi}}, \bibinfo {author} {\bibfnamefont {A.}~\bibnamefont {Cavagna}},
  \bibinfo {author} {\bibfnamefont {L.~D.}\ \bibnamefont {Castello}}, \bibinfo
  {author} {\bibfnamefont {I.}~\bibnamefont {Giardina}}, \bibinfo {author}
  {\bibfnamefont {T.~S.}\ \bibnamefont {Grigera}}, \bibinfo {author}
  {\bibfnamefont {A.}~\bibnamefont {Jeli{\'{c}}}}, \bibinfo {author}
  {\bibfnamefont {S.}~\bibnamefont {Melillo}}, \bibinfo {author} {\bibfnamefont
  {L.}~\bibnamefont {Parisi}}, \bibinfo {author} {\bibfnamefont
  {O.}~\bibnamefont {Pohl}}, \bibinfo {author} {\bibfnamefont {E.}~\bibnamefont
  {Shen}},\ and\ \bibinfo {author} {\bibfnamefont {M.}~\bibnamefont {Viale}},\
  }\bibfield  {title} {\bibinfo {title} {Information transfer and behavioural
  inertia in starling flocks},\ }\href {https://doi.org/10.1038/nphys3035}
  {\bibfield  {journal} {\bibinfo  {journal} {Nature Phys.}\ }\textbf {\bibinfo
  {volume} {10}},\ \bibinfo {pages} {691} (\bibinfo {year} {2014})}\BibitemShut
  {NoStop}%
\bibitem [{\citenamefont {Cavagna}\ \emph
  {et~al.}(2015{\natexlab{a}})\citenamefont {Cavagna}, \citenamefont
  {Castello}, \citenamefont {Giardina}, \citenamefont {Grigera}, \citenamefont
  {Jelic}, \citenamefont {Melillo}, \citenamefont {Mora},\ and\ \citenamefont
  {et~al.}}]{flocking_and_turning}%
  \BibitemOpen
  \bibfield  {author} {\bibinfo {author} {\bibfnamefont {A.}~\bibnamefont
  {Cavagna}}, \bibinfo {author} {\bibfnamefont {L.~D.}\ \bibnamefont
  {Castello}}, \bibinfo {author} {\bibfnamefont {I.}~\bibnamefont {Giardina}},
  \bibinfo {author} {\bibfnamefont {T.}~\bibnamefont {Grigera}}, \bibinfo
  {author} {\bibfnamefont {A.}~\bibnamefont {Jelic}}, \bibinfo {author}
  {\bibfnamefont {S.}~\bibnamefont {Melillo}}, \bibinfo {author} {\bibfnamefont
  {T.}~\bibnamefont {Mora}},\ and\ \bibinfo {author} {\bibfnamefont {L.~P.}\
  \bibnamefont {et~al.}},\ }\bibfield  {title} {\bibinfo {title} {Flocking and
  turning: a new model for self-organized collective motion},\ }\href
  {https://doi.org/10.1007/s10955-014-1119-3} {\bibfield  {journal} {\bibinfo
  {journal} {J. Stat. Phys.}\ }\textbf {\bibinfo {volume} {158}} (\bibinfo
  {year} {2015}{\natexlab{a}})}\BibitemShut {NoStop}%
\bibitem [{\citenamefont {Cavagna}\ \emph
  {et~al.}(2015{\natexlab{b}})\citenamefont {Cavagna}, \citenamefont
  {Giardina}, \citenamefont {Grigera}, \citenamefont {Jelic}, \citenamefont
  {Levine}, \citenamefont {Ramaswamy},\ and\ \citenamefont
  {Viale}}]{cavagna2015silent}%
  \BibitemOpen
  \bibfield  {author} {\bibinfo {author} {\bibfnamefont {A.}~\bibnamefont
  {Cavagna}}, \bibinfo {author} {\bibfnamefont {I.}~\bibnamefont {Giardina}},
  \bibinfo {author} {\bibfnamefont {T.~S.}\ \bibnamefont {Grigera}}, \bibinfo
  {author} {\bibfnamefont {A.}~\bibnamefont {Jelic}}, \bibinfo {author}
  {\bibfnamefont {D.}~\bibnamefont {Levine}}, \bibinfo {author} {\bibfnamefont
  {S.}~\bibnamefont {Ramaswamy}},\ and\ \bibinfo {author} {\bibfnamefont
  {M.}~\bibnamefont {Viale}},\ }\bibfield  {title} {\bibinfo {title} {Silent
  flocks: Constraints on signal propagation across biological groups},\ }\href
  {https://doi.org/10.1103/PhysRevLett.114.218101} {\bibfield  {journal}
  {\bibinfo  {journal} {Phys. Rev. Lett.}\ }\textbf {\bibinfo {volume} {114}},\
  \bibinfo {pages} {218101} (\bibinfo {year} {2015}{\natexlab{b}})}\BibitemShut
  {NoStop}%
\bibitem [{\citenamefont {Vicsek}\ \emph {et~al.}(1995)\citenamefont {Vicsek},
  \citenamefont {Czir\'ok}, \citenamefont {Ben-Jacob}, \citenamefont {Cohen},\
  and\ \citenamefont {Shochet}}]{vicsek}%
  \BibitemOpen
  \bibfield  {author} {\bibinfo {author} {\bibfnamefont {T.}~\bibnamefont
  {Vicsek}}, \bibinfo {author} {\bibfnamefont {A.}~\bibnamefont {Czir\'ok}},
  \bibinfo {author} {\bibfnamefont {E.}~\bibnamefont {Ben-Jacob}}, \bibinfo
  {author} {\bibfnamefont {I.}~\bibnamefont {Cohen}},\ and\ \bibinfo {author}
  {\bibfnamefont {O.}~\bibnamefont {Shochet}},\ }\bibfield  {title} {\bibinfo
  {title} {Novel type of phase transition in a system of self-driven
  particles},\ }\href {https://doi.org/10.1103/PhysRevLett.75.1226} {\bibfield
  {journal} {\bibinfo  {journal} {Phys. Rev. Lett.}\ }\textbf {\bibinfo
  {volume} {75}},\ \bibinfo {pages} {1226} (\bibinfo {year}
  {1995})}\BibitemShut {NoStop}%
\bibitem [{\citenamefont {Vicsek}\ and\ \citenamefont
  {Zafeiris}(2012)}]{Vicsek_review}%
  \BibitemOpen
  \bibfield  {author} {\bibinfo {author} {\bibfnamefont {T.}~\bibnamefont
  {Vicsek}}\ and\ \bibinfo {author} {\bibfnamefont {A.}~\bibnamefont
  {Zafeiris}},\ }\bibfield  {title} {\bibinfo {title} {Collective motion},\
  }\href {https://doi.org/10.1016/j.physrep.2012.03.004} {\bibfield  {journal}
  {\bibinfo  {journal} {Phys. Rep.}\ }\textbf {\bibinfo {volume} {517}},\
  \bibinfo {pages} {71} (\bibinfo {year} {2012})}\BibitemShut {NoStop}%
\bibitem [{\citenamefont {Couzin}\ and\ \citenamefont
  {Krause}(2003)}]{couzin_review}%
  \BibitemOpen
  \bibfield  {author} {\bibinfo {author} {\bibfnamefont {I.~D.}\ \bibnamefont
  {Couzin}}\ and\ \bibinfo {author} {\bibfnamefont {J.}~\bibnamefont
  {Krause}},\ }\bibfield  {title} {\bibinfo {title} {Self-organization and
  collective behavior in vertebrates},\ }\href
  {https://doi.org/10.1016/S0065-3454(03)01001-5} {\bibfield  {journal}
  {\bibinfo  {journal} {Adv. Stud. Behav.}\ }\textbf {\bibinfo {volume} {32}},\
  \bibinfo {pages} {1} (\bibinfo {year} {2003})}\BibitemShut {NoStop}%
\bibitem [{\citenamefont {Toner}\ \emph {et~al.}(2005)\citenamefont {Toner},
  \citenamefont {Tu},\ and\ \citenamefont {Ramaswamy}}]{toner_review}%
  \BibitemOpen
  \bibfield  {author} {\bibinfo {author} {\bibfnamefont {J.}~\bibnamefont
  {Toner}}, \bibinfo {author} {\bibfnamefont {Y.}~\bibnamefont {Tu}},\ and\
  \bibinfo {author} {\bibfnamefont {S.}~\bibnamefont {Ramaswamy}},\ }\bibfield
  {title} {\bibinfo {title} {Hydrodynamics and phases of flocks},\ }\href
  {https://doi.org/10.1016/j.aop.2005.04.011} {\bibfield  {journal} {\bibinfo
  {journal} {Ann. Phys.}\ }\textbf {\bibinfo {volume} {318}},\ \bibinfo {pages}
  {170} (\bibinfo {year} {2005})}\BibitemShut {NoStop}%
\bibitem [{\citenamefont {Marchetti}\ \emph {et~al.}(2013)\citenamefont
  {Marchetti}, \citenamefont {Joanny}, \citenamefont {Ramaswamy}, \citenamefont
  {Liverpool}, \citenamefont {Prost}, \citenamefont {Rao},\ and\ \citenamefont
  {Simha}}]{marchetti_review}%
  \BibitemOpen
  \bibfield  {author} {\bibinfo {author} {\bibfnamefont {M.~C.}\ \bibnamefont
  {Marchetti}}, \bibinfo {author} {\bibfnamefont {J.~F.}\ \bibnamefont
  {Joanny}}, \bibinfo {author} {\bibfnamefont {S.}~\bibnamefont {Ramaswamy}},
  \bibinfo {author} {\bibfnamefont {T.~B.}\ \bibnamefont {Liverpool}}, \bibinfo
  {author} {\bibfnamefont {J.}~\bibnamefont {Prost}}, \bibinfo {author}
  {\bibfnamefont {M.}~\bibnamefont {Rao}},\ and\ \bibinfo {author}
  {\bibfnamefont {R.~A.}\ \bibnamefont {Simha}},\ }\bibfield  {title} {\bibinfo
  {title} {Hydrodynamics of soft active matter},\ }\href
  {https://doi.org/10.1103/RevModPhys.85.1143} {\bibfield  {journal} {\bibinfo
  {journal} {Rev. Mod. Phys.}\ }\textbf {\bibinfo {volume} {85}},\ \bibinfo
  {pages} {1143} (\bibinfo {year} {2013})}\BibitemShut {NoStop}%
\bibitem [{\citenamefont {Ramaswamy}(2010)}]{ramaswamy_review}%
  \BibitemOpen
  \bibfield  {author} {\bibinfo {author} {\bibfnamefont {S.}~\bibnamefont
  {Ramaswamy}},\ }\bibfield  {title} {\bibinfo {title} {The mechanics and
  statistics of active matter},\ }\href
  {https://doi.org/10.1146/annurev-conmatphys-070909-104101} {\bibfield
  {journal} {\bibinfo  {journal} {Annu. Rev. Conden. Ma. P.}\ }\textbf
  {\bibinfo {volume} {1}},\ \bibinfo {pages} {323} (\bibinfo {year}
  {2010})}\BibitemShut {NoStop}%
\bibitem [{\citenamefont {Cavagna}\ \emph {et~al.}(2018)\citenamefont
  {Cavagna}, \citenamefont {Giardina},\ and\ \citenamefont
  {Grigera}}]{Review_2018}%
  \BibitemOpen
  \bibfield  {author} {\bibinfo {author} {\bibfnamefont {A.}~\bibnamefont
  {Cavagna}}, \bibinfo {author} {\bibfnamefont {I.}~\bibnamefont {Giardina}},\
  and\ \bibinfo {author} {\bibfnamefont {T.~S.}\ \bibnamefont {Grigera}},\
  }\bibfield  {title} {\bibinfo {title} {The physics of flocking: Correlation
  as a compass from experiments to theory},\ }\href
  {https://doi.org/https://doi.org/10.1016/j.physrep.2017.11.003} {\bibfield
  {journal} {\bibinfo  {journal} {Phys. Rep.}\ }\textbf {\bibinfo {volume}
  {728}},\ \bibinfo {pages} {1} (\bibinfo {year} {2018})}\BibitemShut {NoStop}%
\bibitem [{\citenamefont {Zwanzig}(2001)}]{zwanzig_book}%
  \BibitemOpen
  \bibfield  {author} {\bibinfo {author} {\bibfnamefont {R.}~\bibnamefont
  {Zwanzig}},\ }\href@noop {} {\emph {\bibinfo {title} {Nonequilibrium
  statistical mechanics}}}\ (\bibinfo  {publisher} {Oxford University Press},\
  \bibinfo {year} {2001})\BibitemShut {NoStop}%
\bibitem [{\citenamefont {Attanasi}\ \emph {et~al.}(2015)\citenamefont
  {Attanasi}, \citenamefont {Cavagna}, \citenamefont {Castello}, \citenamefont
  {Giardina}, \citenamefont {Jelic}, \citenamefont {Melillo}, \citenamefont
  {Parisi}, \citenamefont {Pohl}, \citenamefont {Shen},\ and\ \citenamefont
  {Viale}}]{attanasi2015emergence}%
  \BibitemOpen
  \bibfield  {author} {\bibinfo {author} {\bibfnamefont {A.}~\bibnamefont
  {Attanasi}}, \bibinfo {author} {\bibfnamefont {A.}~\bibnamefont {Cavagna}},
  \bibinfo {author} {\bibfnamefont {L.~D.}\ \bibnamefont {Castello}}, \bibinfo
  {author} {\bibfnamefont {I.}~\bibnamefont {Giardina}}, \bibinfo {author}
  {\bibfnamefont {A.}~\bibnamefont {Jelic}}, \bibinfo {author} {\bibfnamefont
  {S.}~\bibnamefont {Melillo}}, \bibinfo {author} {\bibfnamefont
  {L.}~\bibnamefont {Parisi}}, \bibinfo {author} {\bibfnamefont
  {O.}~\bibnamefont {Pohl}}, \bibinfo {author} {\bibfnamefont {E.}~\bibnamefont
  {Shen}},\ and\ \bibinfo {author} {\bibfnamefont {M.}~\bibnamefont {Viale}},\
  }\bibfield  {title} {\bibinfo {title} {Emergence of collective changes in
  travel direction of starling flocks from individual birds{\textquotesingle}
  fluctuations},\ }\href {https://doi.org/10.1098/rsif.2015.0319} {\bibfield
  {journal} {\bibinfo  {journal} {J. R. Soc. Interface}\ }\textbf {\bibinfo
  {volume} {12}},\ \bibinfo {pages} {20150319} (\bibinfo {year}
  {2015})}\BibitemShut {NoStop}%
\bibitem [{\citenamefont {Cavagna}\ \emph {et~al.}(2017)\citenamefont
  {Cavagna}, \citenamefont {Giardina}, \citenamefont {Jelic}, \citenamefont
  {Melillo}, \citenamefont {Parisi}, \citenamefont {Silvestri},\ and\
  \citenamefont {Viale}}]{Cavagna_2017}%
  \BibitemOpen
  \bibfield  {author} {\bibinfo {author} {\bibfnamefont {A.}~\bibnamefont
  {Cavagna}}, \bibinfo {author} {\bibfnamefont {I.}~\bibnamefont {Giardina}},
  \bibinfo {author} {\bibfnamefont {A.}~\bibnamefont {Jelic}}, \bibinfo
  {author} {\bibfnamefont {S.}~\bibnamefont {Melillo}}, \bibinfo {author}
  {\bibfnamefont {L.}~\bibnamefont {Parisi}}, \bibinfo {author} {\bibfnamefont
  {E.}~\bibnamefont {Silvestri}},\ and\ \bibinfo {author} {\bibfnamefont
  {M.}~\bibnamefont {Viale}},\ }\bibfield  {title} {\bibinfo {title}
  {Nonsymmetric interactions trigger collective swings in globally ordered
  systems},\ }\href {https://doi.org/10.1103/PhysRevLett.118.138003} {\bibfield
   {journal} {\bibinfo  {journal} {Phys. Rev. Lett.}\ }\textbf {\bibinfo
  {volume} {118}},\ \bibinfo {pages} {138003} (\bibinfo {year}
  {2017})}\BibitemShut {NoStop}%
\bibitem [{\citenamefont {Venturelli}(2019)}]{tesi_venturelli}%
  \BibitemOpen
  \bibfield  {author} {\bibinfo {author} {\bibfnamefont {D.}~\bibnamefont
  {Venturelli}},\ }\emph {\bibinfo {title} {Dynamical response to local
  perturbations of an active matter system with polar order}},\ \href
  {https://doi.org/10.5281/zenodo.7615248} {Master's thesis},\ \bibinfo
  {school} {{Sapienza, Universit\a`{a} di Roma}} (\bibinfo {year}
  {2019})\BibitemShut {NoStop}%
\bibitem [{\citenamefont {Mora}\ \emph {et~al.}(2016)\citenamefont {Mora},
  \citenamefont {Walczak}, \citenamefont {Castello}, \citenamefont {Ginelli},
  \citenamefont {Melillo}, \citenamefont {Parisi}, \citenamefont {Viale},
  \citenamefont {Cavagna},\ and\ \citenamefont {Giardina}}]{Nature_2016}%
  \BibitemOpen
  \bibfield  {author} {\bibinfo {author} {\bibfnamefont {T.}~\bibnamefont
  {Mora}}, \bibinfo {author} {\bibfnamefont {A.~M.}\ \bibnamefont {Walczak}},
  \bibinfo {author} {\bibfnamefont {L.~D.}\ \bibnamefont {Castello}}, \bibinfo
  {author} {\bibfnamefont {F.}~\bibnamefont {Ginelli}}, \bibinfo {author}
  {\bibfnamefont {S.}~\bibnamefont {Melillo}}, \bibinfo {author} {\bibfnamefont
  {L.}~\bibnamefont {Parisi}}, \bibinfo {author} {\bibfnamefont
  {M.}~\bibnamefont {Viale}}, \bibinfo {author} {\bibfnamefont
  {A.}~\bibnamefont {Cavagna}},\ and\ \bibinfo {author} {\bibfnamefont
  {I.}~\bibnamefont {Giardina}},\ }\bibfield  {title} {\bibinfo {title} {Local
  equilibrium in bird flocks},\ }\href {https://doi.org/10.1038/nphys3846}
  {\bibfield  {journal} {\bibinfo  {journal} {Nature Phys.}\ }\textbf {\bibinfo
  {volume} {12}},\ \bibinfo {pages} {1153} (\bibinfo {year}
  {2016})}\BibitemShut {NoStop}%
\bibitem [{\citenamefont {Vanden-Eijnden}\ and\ \citenamefont
  {Ciccotti}(2006)}]{ciccotti}%
  \BibitemOpen
  \bibfield  {author} {\bibinfo {author} {\bibfnamefont {E.}~\bibnamefont
  {Vanden-Eijnden}}\ and\ \bibinfo {author} {\bibfnamefont {G.}~\bibnamefont
  {Ciccotti}},\ }\bibfield  {title} {\bibinfo {title} {Second-order integrators
  for {Langevin} equations with holonomic constraints},\ }\href
  {https://doi.org/https://doi.org/10.1016/j.cplett.2006.07.086} {\bibfield
  {journal} {\bibinfo  {journal} {Chem. Phys. Lett.}\ }\textbf {\bibinfo
  {volume} {429}},\ \bibinfo {pages} {310} (\bibinfo {year}
  {2006})}\BibitemShut {NoStop}%
\bibitem [{\citenamefont {Chepizhko}\ \emph {et~al.}(2021)\citenamefont
  {Chepizhko}, \citenamefont {Saintillan},\ and\ \citenamefont
  {Peruani}}]{chepizhko2021revisiting}%
  \BibitemOpen
  \bibfield  {author} {\bibinfo {author} {\bibfnamefont {O.}~\bibnamefont
  {Chepizhko}}, \bibinfo {author} {\bibfnamefont {D.}~\bibnamefont
  {Saintillan}},\ and\ \bibinfo {author} {\bibfnamefont {F.}~\bibnamefont
  {Peruani}},\ }\bibfield  {title} {\bibinfo {title} {Revisiting the emergence
  of order in active matter},\ }\href {https://doi.org/10.1039/D0SM01220C}
  {\bibfield  {journal} {\bibinfo  {journal} {Soft Matter}\ }\textbf {\bibinfo
  {volume} {17}},\ \bibinfo {pages} {3113} (\bibinfo {year}
  {2021})}\BibitemShut {NoStop}%
\bibitem [{\citenamefont {Frenkel}\ and\ \citenamefont {Smit}(2001)}]{frenkel}%
  \BibitemOpen
  \bibfield  {author} {\bibinfo {author} {\bibfnamefont {D.}~\bibnamefont
  {Frenkel}}\ and\ \bibinfo {author} {\bibfnamefont {B.}~\bibnamefont {Smit}},\
  }\href@noop {} {\emph {\bibinfo {title} {Understanding Molecular
  Simulation}}},\ \bibinfo {edition} {2nd}\ ed.\ (\bibinfo  {publisher}
  {Academic Press, Inc.},\ \bibinfo {address} {USA},\ \bibinfo {year}
  {2001})\BibitemShut {NoStop}%
\bibitem [{\citenamefont {Tuckerman}(2010)}]{tuckerman}%
  \BibitemOpen
  \bibfield  {author} {\bibinfo {author} {\bibfnamefont {M.~E.}\ \bibnamefont
  {Tuckerman}},\ }\href@noop {} {\emph {\bibinfo {title} {Statistical
  Mechanics}}}\ (\bibinfo  {publisher} {Oxford University Press},\ \bibinfo
  {year} {2010})\BibitemShut {NoStop}%
\bibitem [{\citenamefont {Venturelli}\ and\ \citenamefont
  {Loffredo}(2022)}]{code}%
  \BibitemOpen
  \bibfield  {author} {\bibinfo {author} {\bibfnamefont {D.}~\bibnamefont
  {Venturelli}}\ and\ \bibinfo {author} {\bibfnamefont {E.}~\bibnamefont
  {Loffredo}},\ }\href {https://github.com/sonarventu/ISM} {\bibinfo {title}
  {{Source code for the 2D Inertial Spin Model}}} (\bibinfo {year}
  {2022})\BibitemShut {NoStop}%
\bibitem [{\citenamefont {Cavagna}\ \emph
  {et~al.}(2015{\natexlab{c}})\citenamefont {Cavagna}, \citenamefont
  {Del~Castello}, \citenamefont {Dey}, \citenamefont {Giardina}, \citenamefont
  {Melillo}, \citenamefont {Parisi},\ and\ \citenamefont
  {Viale}}]{cavagna2015short}%
  \BibitemOpen
  \bibfield  {author} {\bibinfo {author} {\bibfnamefont {A.}~\bibnamefont
  {Cavagna}}, \bibinfo {author} {\bibfnamefont {L.}~\bibnamefont
  {Del~Castello}}, \bibinfo {author} {\bibfnamefont {S.}~\bibnamefont {Dey}},
  \bibinfo {author} {\bibfnamefont {I.}~\bibnamefont {Giardina}}, \bibinfo
  {author} {\bibfnamefont {S.}~\bibnamefont {Melillo}}, \bibinfo {author}
  {\bibfnamefont {L.}~\bibnamefont {Parisi}},\ and\ \bibinfo {author}
  {\bibfnamefont {M.}~\bibnamefont {Viale}},\ }\bibfield  {title} {\bibinfo
  {title} {Short-range interactions versus long-range correlations in bird
  flocks},\ }\href {https://doi.org/10.1103/PhysRevE.92.012705} {\bibfield
  {journal} {\bibinfo  {journal} {Phys. Rev. E}\ }\textbf {\bibinfo {volume}
  {92}},\ \bibinfo {pages} {012705} (\bibinfo {year}
  {2015}{\natexlab{c}})}\BibitemShut {NoStop}%
\bibitem [{\citenamefont {Ferretti}\ \emph {et~al.}(2020)\citenamefont
  {Ferretti}, \citenamefont {Chard\`es}, \citenamefont {Mora}, \citenamefont
  {Walczak},\ and\ \citenamefont {Giardina}}]{ferretti2020building}%
  \BibitemOpen
  \bibfield  {author} {\bibinfo {author} {\bibfnamefont {F.}~\bibnamefont
  {Ferretti}}, \bibinfo {author} {\bibfnamefont {V.}~\bibnamefont {Chard\`es}},
  \bibinfo {author} {\bibfnamefont {T.}~\bibnamefont {Mora}}, \bibinfo {author}
  {\bibfnamefont {A.~M.}\ \bibnamefont {Walczak}},\ and\ \bibinfo {author}
  {\bibfnamefont {I.}~\bibnamefont {Giardina}},\ }\bibfield  {title} {\bibinfo
  {title} {Building general {Langevin} models from discrete datasets},\ }\href
  {https://doi.org/10.1103/PhysRevX.10.031018} {\bibfield  {journal} {\bibinfo
  {journal} {Phys. Rev. X}\ }\textbf {\bibinfo {volume} {10}},\ \bibinfo
  {pages} {031018} (\bibinfo {year} {2020})}\BibitemShut {NoStop}%
\bibitem [{\citenamefont {Cavagna}\ \emph {et~al.}(2021)\citenamefont
  {Cavagna}, \citenamefont {Di~Carlo}, \citenamefont {Giardina}, \citenamefont
  {Grigera}, \citenamefont {Melillo}, \citenamefont {Parisi}, \citenamefont
  {Pisegna},\ and\ \citenamefont {Scandolo}}]{3.99}%
  \BibitemOpen
  \bibfield  {author} {\bibinfo {author} {\bibfnamefont {A.}~\bibnamefont
  {Cavagna}}, \bibinfo {author} {\bibfnamefont {L.}~\bibnamefont {Di~Carlo}},
  \bibinfo {author} {\bibfnamefont {I.}~\bibnamefont {Giardina}}, \bibinfo
  {author} {\bibfnamefont {T.~S.}\ \bibnamefont {Grigera}}, \bibinfo {author}
  {\bibfnamefont {S.}~\bibnamefont {Melillo}}, \bibinfo {author} {\bibfnamefont
  {L.}~\bibnamefont {Parisi}}, \bibinfo {author} {\bibfnamefont
  {G.}~\bibnamefont {Pisegna}},\ and\ \bibinfo {author} {\bibfnamefont
  {M.}~\bibnamefont {Scandolo}},\ }\href
  {https://doi.org/10.48550/ARXIV.2107.04432} {\bibinfo {title} {Natural swarms
  in $\bf 3.99$ dimensions}} (\bibinfo {year} {2021})\BibitemShut {NoStop}%
\bibitem [{\citenamefont {Holubec}(2023)}]{private}%
  \BibitemOpen
  \bibfield  {author} {\bibinfo {author} {\bibfnamefont {V.}~\bibnamefont
  {Holubec}},\ }\href@noop {} {}\bibinfo {howpublished} {private communication}
  (\bibinfo {year} {2023})\BibitemShut {NoStop}%
\bibitem [{\citenamefont {Sakurai}\ and\ \citenamefont
  {Napolitano}(2017)}]{sakurai2017modern}%
  \BibitemOpen
  \bibfield  {author} {\bibinfo {author} {\bibfnamefont {J.}~\bibnamefont
  {Sakurai}}\ and\ \bibinfo {author} {\bibfnamefont {J.}~\bibnamefont
  {Napolitano}},\ }\href {https://books.google.it/books?id=010yDwAAQBAJ} {\emph
  {\bibinfo {title} {Modern Quantum Mechanics}}}\ (\bibinfo  {publisher}
  {Cambridge University Press},\ \bibinfo {year} {2017})\BibitemShut {NoStop}%
\bibitem [{\citenamefont {Leoncini}\ \emph {et~al.}(1998)\citenamefont
  {Leoncini}, \citenamefont {Verga},\ and\ \citenamefont
  {Ruffo}}]{Leoncini_98}%
  \BibitemOpen
  \bibfield  {author} {\bibinfo {author} {\bibfnamefont {X.}~\bibnamefont
  {Leoncini}}, \bibinfo {author} {\bibfnamefont {A.~D.}\ \bibnamefont
  {Verga}},\ and\ \bibinfo {author} {\bibfnamefont {S.}~\bibnamefont {Ruffo}},\
  }\bibfield  {title} {\bibinfo {title} {Hamiltonian dynamics and the phase
  transition of the $\mathrm{XY}$ model},\ }\href
  {https://doi.org/10.1103/PhysRevE.57.6377} {\bibfield  {journal} {\bibinfo
  {journal} {Phys. Rev. E}\ }\textbf {\bibinfo {volume} {57}},\ \bibinfo
  {pages} {6377} (\bibinfo {year} {1998})}\BibitemShut {NoStop}%
\bibitem [{\citenamefont {Lepri}\ and\ \citenamefont
  {Ruffo}(2001)}]{Lepri_2001}%
  \BibitemOpen
  \bibfield  {author} {\bibinfo {author} {\bibfnamefont {S.}~\bibnamefont
  {Lepri}}\ and\ \bibinfo {author} {\bibfnamefont {S.}~\bibnamefont {Ruffo}},\
  }\bibfield  {title} {\bibinfo {title} {{Finite-size effects on the
  Hamiltonian dynamics of the XY-model}},\ }\href
  {https://doi.org/10.1209/epl/i2001-00445-5} {\bibfield  {journal} {\bibinfo
  {journal} {Europhys. Lett.}\ }\textbf {\bibinfo {volume} {55}},\ \bibinfo
  {pages} {512} (\bibinfo {year} {2001})}\BibitemShut {NoStop}%
\bibitem [{\citenamefont {Brambati}\ \emph {et~al.}(2022)\citenamefont
  {Brambati}, \citenamefont {Fava},\ and\ \citenamefont
  {Ginelli}}]{Ginelli_2022}%
  \BibitemOpen
  \bibfield  {author} {\bibinfo {author} {\bibfnamefont {M.}~\bibnamefont
  {Brambati}}, \bibinfo {author} {\bibfnamefont {G.}~\bibnamefont {Fava}},\
  and\ \bibinfo {author} {\bibfnamefont {F.}~\bibnamefont {Ginelli}},\
  }\bibfield  {title} {\bibinfo {title} {Signatures of directed and spontaneous
  flocking},\ }\href {https://doi.org/10.1103/PhysRevE.106.024608} {\bibfield
  {journal} {\bibinfo  {journal} {Phys. Rev. E}\ }\textbf {\bibinfo {volume}
  {106}},\ \bibinfo {pages} {024608} (\bibinfo {year} {2022})}\BibitemShut
  {NoStop}%
\bibitem [{\citenamefont {Le~Bellac}(1991)}]{LeBellac}%
  \BibitemOpen
  \bibfield  {author} {\bibinfo {author} {\bibfnamefont {M.}~\bibnamefont
  {Le~Bellac}},\ }\href@noop {} {\emph {\bibinfo {title} {Quantum and
  statistical field theory}}}\ (\bibinfo  {publisher} {{Clarendon Press}},\
  \bibinfo {year} {1991})\BibitemShut {NoStop}%
\bibitem [{\citenamefont {Mermin}\ and\ \citenamefont
  {Wagner}(1966)}]{merminwagner}%
  \BibitemOpen
  \bibfield  {author} {\bibinfo {author} {\bibfnamefont {N.~D.}\ \bibnamefont
  {Mermin}}\ and\ \bibinfo {author} {\bibfnamefont {H.}~\bibnamefont
  {Wagner}},\ }\bibfield  {title} {\bibinfo {title} {Absence of ferromagnetism
  or antiferromagnetism in one- or two-dimensional isotropic {Heisenberg}
  models},\ }\href {https://doi.org/10.1103/PhysRevLett.17.1133} {\bibfield
  {journal} {\bibinfo  {journal} {Phys. Rev. Lett.}\ }\textbf {\bibinfo
  {volume} {17}},\ \bibinfo {pages} {1133} (\bibinfo {year}
  {1966})}\BibitemShut {NoStop}%
\bibitem [{\citenamefont {Kosterlitz}\ and\ \citenamefont
  {Thouless}(1973)}]{Kosterlitz_1973}%
  \BibitemOpen
  \bibfield  {author} {\bibinfo {author} {\bibfnamefont {J.~M.}\ \bibnamefont
  {Kosterlitz}}\ and\ \bibinfo {author} {\bibfnamefont {D.~J.}\ \bibnamefont
  {Thouless}},\ }\bibfield  {title} {\bibinfo {title} {Ordering, metastability
  and phase transitions in two-dimensional systems},\ }\href
  {https://doi.org/10.1088/0022-3719/6/7/010} {\bibfield  {journal} {\bibinfo
  {journal} {J. Phys. C}\ }\textbf {\bibinfo {volume} {6}},\ \bibinfo {pages}
  {1181} (\bibinfo {year} {1973})}\BibitemShut {NoStop}%
\bibitem [{\citenamefont {Bramwell}\ and\ \citenamefont
  {Holdsworth}(1993)}]{Bramwell_1993}%
  \BibitemOpen
  \bibfield  {author} {\bibinfo {author} {\bibfnamefont {S.~T.}\ \bibnamefont
  {Bramwell}}\ and\ \bibinfo {author} {\bibfnamefont {P.~C.~W.}\ \bibnamefont
  {Holdsworth}},\ }\bibfield  {title} {\bibinfo {title} {Magnetization and
  universal sub-critical behaviour in two-dimensional {XY} magnets},\ }\href
  {https://doi.org/10.1088/0953-8984/5/4/004} {\bibfield  {journal} {\bibinfo
  {journal} {J. Phys. -- Condens. Mat.}\ }\textbf {\bibinfo {volume} {5}},\
  \bibinfo {pages} {L53} (\bibinfo {year} {1993})}\BibitemShut {NoStop}%
\bibitem [{\citenamefont {Roomany}\ and\ \citenamefont
  {Wyld}(1980)}]{Roomany_80}%
  \BibitemOpen
  \bibfield  {author} {\bibinfo {author} {\bibfnamefont {H.~H.}\ \bibnamefont
  {Roomany}}\ and\ \bibinfo {author} {\bibfnamefont {H.~W.}\ \bibnamefont
  {Wyld}},\ }\bibfield  {title} {\bibinfo {title} {{Finite-lattice approach to
  the O(2) and O(3) models in 1 + 1 dimensions and the (2+1)-dimensional Ising
  model}},\ }\href {https://doi.org/10.1103/PhysRevD.21.3341} {\bibfield
  {journal} {\bibinfo  {journal} {Phys. Rev. D}\ }\textbf {\bibinfo {volume}
  {21}},\ \bibinfo {pages} {3341} (\bibinfo {year} {1980})}\BibitemShut
  {NoStop}%
\bibitem [{\citenamefont {Chung}(1999)}]{Chung_99}%
  \BibitemOpen
  \bibfield  {author} {\bibinfo {author} {\bibfnamefont {S.~G.}\ \bibnamefont
  {Chung}},\ }\bibfield  {title} {\bibinfo {title} {{Essential finite-size
  effect in the two-dimensional XY model}},\ }\href
  {https://doi.org/10.1103/PhysRevB.60.11761} {\bibfield  {journal} {\bibinfo
  {journal} {Phys. Rev. B}\ }\textbf {\bibinfo {volume} {60}},\ \bibinfo
  {pages} {11761} (\bibinfo {year} {1999})}\BibitemShut {NoStop}%
\bibitem [{\citenamefont {Tobochnik}\ and\ \citenamefont
  {Chester}(1979)}]{Chester_1979}%
  \BibitemOpen
  \bibfield  {author} {\bibinfo {author} {\bibfnamefont {J.}~\bibnamefont
  {Tobochnik}}\ and\ \bibinfo {author} {\bibfnamefont {G.~V.}\ \bibnamefont
  {Chester}},\ }\bibfield  {title} {\bibinfo {title} {{Monte Carlo study of the
  planar spin model}},\ }\href {https://doi.org/10.1103/PhysRevB.20.3761}
  {\bibfield  {journal} {\bibinfo  {journal} {Phys. Rev. B}\ }\textbf {\bibinfo
  {volume} {20}},\ \bibinfo {pages} {3761} (\bibinfo {year}
  {1979})}\BibitemShut {NoStop}%
\bibitem [{\citenamefont {Toner}\ and\ \citenamefont {Tu}(1998)}]{Toner_1998}%
  \BibitemOpen
  \bibfield  {author} {\bibinfo {author} {\bibfnamefont {J.}~\bibnamefont
  {Toner}}\ and\ \bibinfo {author} {\bibfnamefont {Y.}~\bibnamefont {Tu}},\
  }\bibfield  {title} {\bibinfo {title} {Flocks, herds, and schools: A
  quantitative theory of flocking},\ }\href
  {https://doi.org/10.1103/PhysRevE.58.4828} {\bibfield  {journal} {\bibinfo
  {journal} {Phys. Rev. E}\ }\textbf {\bibinfo {volume} {58}},\ \bibinfo
  {pages} {4828} (\bibinfo {year} {1998})}\BibitemShut {NoStop}%
\end{thebibliography}%

\end{document}